\documentclass[%
 reprint,
 amsmath,amssymb,
 aps,prl,
floatfix
]{revtex4-2}

\usepackage{amsmath}
\usepackage{mathtools}
\usepackage{xcolor}
\usepackage{graphicx}% Include figure files
\usepackage{dcolumn}% Align table columns on decimal point
\usepackage{bm}% bold math
\usepackage{float}

\begin{document}
\preprint{APS/123-QED}
\title{Going Beyond the Cumulant Approximation:\\Power Series Correction to Single Particle Green's Function in Holstein System }
\author{Bipul Pandey}
\email{bipulpandey2004@uchicago.edu}
\author{Peter B. Littlewood}
\affiliation{Department of Physics, University of Chicago,Chicago, Illinois, 60637, USA}

\date{\today}% It is always \today, today,
             %  but any date may be explicitly specified

\begin{abstract}
In the context of a single electron two orbital Holstein system coupled to dispersionless bosons, we develop a general method to correct single particle Green’s function using a power series correction (PSC) scheme. We then outline the derivations of various flavors of cumulant approximation through the PSC scheme and explain the assumptions and approximations behind them. Finally, we compute and compare PSC spectral function with cumulant and exact diagonalized spectral functions and elucidate three regimes of this problem - two that cumulant explains and one where cumulant fails. We find that the exact and the PSC spectral functions match within spectral broadening across all three regimes.
\end{abstract}

%\keywords{Suggested keywords}%Use showkeys class option if keyword
                              %display desired
\maketitle

%\tableofcontents

\section{I. Overview:\protect}
Electrons and holes in materials undergo numerous complex interactions among themselves, the external fields, as well as the constituent atomic lattice. The strength of such many body interactions depends on various factors such as electronic configuration of the host material, presence of doping and defects, lattice parameters etc. Such factors manifest as bosonic collective excitations that renormalize the particle states (electrons/holes) into quasiparticles states with different energy and lifetime, and even mix quasiparticle states depending on the interaction strength. Alongside the quasiparticle features in photo-emission spectra, these collective excitations show up as "shake-off" features that can be loosely separated into sharp satellites emerging from bosonic collective modes (such as plasmons and optical phonons) and continua arising from non-zero-momentum particle-hole excitations (including excitons) \cite{goulielmakis_real-time_2010,lemell_real-time_2015,neppl_direct_2015}.

In calculations, the interaction strengths between collective excitations and particles are modeled as tunable electron-boson coupling parameters. In experiments, this coupling tunability is achieved by introducing doping and defects \cite{wang_tailoring_2016, kang_holstein_2018}. Although at very weak coupling the quasiparticle renormalization due to the collective modes is negligible, with stronger coupling a proportional renormalization of the quasiparticle occurs. As an example, in photo-emission spectra of strontium titanate this coupling manifests as a significant shift in quasiparticle energy, significant decrease in lifetime and intensity of quasiparticle features, strong shake-off features, as well as a strong mass enhancement of the carrier \cite{van_mechelen_electron-phonon_2008, devreese_many-body_2010, wang2016, swartz2018, edelman2021}. Strong electron-phonon coupling is also visible in electronic spectra in metallic cuprates \cite{rosch_polaronic_2005,damascelli_angle-resolved_2003} and the metal-insulator transition in undoped cuprates \cite{baldini_electronphonon-driven_2020}, and other correlated metals for example $FeSe/SrTiO_3$ epitaxial layers \cite{yang2015}.
At extreme values of coupling constant, strong electron-boson coupling can completely self-trap and localize electrons creating polaronic states. This severely alter carrier mobility in the material. This is of particular interest in the material design for photovoltaics and electronics \cite{mohamed_electronic_2019, ma_energy_2016, hulea_tunable_2006}. Finally, in the presence of multiple boson species, there can be competition between their effect on the carrier which creates novel phase crossovers in materials \cite{riley_crossover_2018}. Therefore a proper understanding and quantification of the effects of collective modes on charge carriers is vital in understanding and designing novel material with interesting engineering applications.

 In this work, we build on, and generalize, existing non-perturbative methods including the "GW" approximation \cite{hedin_new_1965} and the cumulant expansion \cite{langreth_singularities_1970} to describe the single particle dynamics of a system with multiple electronic levels interacting through common boson baths. The paper is organized as follows. In Section II, we introduce the model problem and the concepts of electron Green's function, electron self energy and cumulant corrections. In III, we briefly introduce the existing methods and their major drawbacks. In IV, we develop our correction scheme, and physically motivate the assumptions used to simplify the equations. In sections V, we outline the derivation of various flavors of cumulants through our method and elucidate the implicitly made but vaguely understood assumptions behind these approximation. Finally, in section VI, we identify three important regimes of the problem by comparing the performance of the cumulant method and the power series method with results from exact diagonalization of this problem in finite boson basis.
 %%%%%%%%%%%%%%%%%%%%%%%%%%%%%%%%%%%%%%%%%%%%%%%%%%%
\section{II. Introduction to the Problem}
 We consider a model Hamiltonian for a single electron two orbital Holstein system with bonding/anti-bonding energy $\varepsilon_+/\varepsilon_-$ such that their difference is $\Delta$. This system is kept in baths of two dispersionless boson species $(\pm)$. The bosons are quantized packets of energy $\omega_o$ that the electron can interact with. Interaction of electron with $(-)$ bosons causes an electron's inter-orbital transition. The $(+)$ boson does not cause any electronic transition upon interaction. 
 
 The electron-boson interaction strength is controlled by a coupling constant $g$. The fermionic ladder operators are $c_+/c_+^\dagger$ and $c_-/c_-^\dagger$ for bonding and anti-bonding orbitals respectively. The bosonic ladder operators for $(\pm$) bosons are $b_\pm /b_\pm^\dagger$. The Hamiltonian for this problem is separable into three distinct pieces. $H_o$ is the non-interacting part of the Hamiltonian. $H_+$ explicitly has $(+)$ bosons and doesn't cause inter-orbital transitions while $H_-$ explicitly has $(-)$ bosons and governs inter-orbital transitions.
\begin{equation}
\begin{aligned}
H\, &= H_o + H_+ + H_- \quad\text{where,} \\
H_o &= \sum_{i=\pm} \varepsilon_\pm c_i^\dagger c_i  \\
H_+ &=  \omega_o b_+^\dagger b_+ +  g(c_+^\dagger c_+ + c_-^\dagger c_-)(b_+^\dagger + b_+) \\
H_- &=  \omega_o b_-^\dagger b_- +g(c_+^\dagger c_- + c_-^\dagger c_+)(b_-^\dagger + b_-)
\end{aligned}
\label {Holstein}
\end{equation}
%%%%%%%%%%%%%%%%%%%%%%%%%%%
%%%%%%%%%%%%%%%%%%%%%%%%%
 Here $g$ is same in both $H_\pm$ due to the original problem's symmetries. But, even if they are different i.e $g_\pm$ in $H_\pm$, we can find corrections in powers of a dummy variable $g$ that multiplies both $g_\pm$ and set it to $1$ in the end.

This Hamiltonian describes the physics of a model of the dihydrogen cation $(H_2^+)$ - two hydrogen nuclei and a single electron. Historically, this problem was approached with clamped nuclei approximation. This crude approach completely neglects the vibronic coupling between the electron and vibrational modes of the nuclei (optical phonons in crystalline structures - see supplement) which becomes crucial when $\Delta \approx \omega_o$. Vibronic couplings in this regime can cause inter-band transitions and severely renormalize the energy levels in the molecule \cite{heller_molecular_1979,heller_molecular_1980, ranninger_two-site_1992}. Hence, this is a good model to construct the approximation scheme due of its simplicity and similarities to real multi-level systems. Furthermore, no exact analytical solution exists and the approximate methods either give incorrect boson satellites (GW) or are {\it ad hoc}, unsystematic and incorrect at strong coupling (cumulant) \cite{gunnarsson_corrections_1994}. 
%%%%%%%%%%%%%%%%%%%%%%%%%%%%%%%%%%%%%%%%
%%%%%%%%%%%%%%%%%%%%%%%%%%%%%%%%%%%%%%%%%

%\subsection{Electron Green's Function and Spectral Function}
\textbf{The Green's function:}
The retarded-time(RT) formalism is better suited to handle electron-hole interactions because it treats both of them in equal footing as particles \cite{kas_cumulant_2014}. For the Holstein problem \eqref{Holstein} with fock vacuum $|0\rangle$ as the ground state and \{ , \}/[ , ] as the anti-commutator/commutator, The electron Green's function $G(n,t)$ for each orbital ($n=\pm$) and the boson Green's function $\mathcal{D}(N,t)$ for each boson species $(N =\pm)$ in retarded formalism is given by;
\begin{equation}
\begin{aligned}
    G(n,t) &= -i \theta(t)\langle 0|\{c_n(t) ,{c_n^{\dagger}}\}|0\rangle\\
    \mathcal{D}(N,t) &= -i \theta(t)\langle 0|[b_N(t) ,{b_N^{\dagger}}]|0\rangle\\
\end{aligned}
\label{Greens definition}
\end{equation}
For non-interacting(g=0) electrons and dispersionless bosons with energy $\omega_o$, the bare electron green's function $G_o$ and a bare boson green's function $\mathcal{D}$ are,
\begin{equation}
\begin{aligned}
G_o(\pm,t) = -i\theta(t) e^{-i\varepsilon_\pm t} \\ \mathcal{D}(\pm,t) = -i\theta(t)e^{-i\omega_o t}
\end{aligned}
\label{Bare electron and boson greens and greens}
\end{equation}

The quasiparticle energies, lifetimes and the boson satellites show up as complex poles of $G(n,\omega)$ where $\omega$ is the frequency. The frequency axis spectral function, $A(m,n;\omega)$ (see supplement) is defined as;
 \begin{equation}
     A(m,n;\omega) = \frac{1}{\pi} |\text{Im}G(m,n;\omega)|
 \end{equation}
%%%%%%%%%%%%%%%%%%%%%%%%%%%%%%%%%%%%%%%%%
%\subsection{Electron Self Energy and Dyson's equation}
\textbf{Electron self energy and Dyson's equation: }
  At zero coupling ($g=0$), the energy eigenvalues $\varepsilon_{\pm}$ of \eqref{Holstein} are real and the states have infinite lifetime owing to the lack of interaction between the orbitals. However, upon switching on the boson mediated interaction ($g\neq0$) between orbitals, the exchange of energy and momenta between states through boson exchange causes clumping of electrons and holes to form quasiparticles. Because of time-transnational invariance, we can package this interaction information together and call it the self energy.
\begin{equation}
    -i\Sigma(t) = g^2\!\sum_{\substack{N=\pm\\n=\pm}} \mathcal{D}(N,t)G(n, t)= g^2 \!\sum_{n = \pm}\!\!\!-i\Sigma(n,t)
\label{self-energy}
\end{equation}
Each orbital's self energy $\Sigma(n, t)$ is complex valued unlike the bare energy. This gives rise to spectral peak broadening - an indicative of finite quasiparticle lifetime. A properly constructed self energy also incorporates boson mediated inter-orbital transitions, produces satellite peaks at the correct boson frequencies and redistributes the spectral weight from the quasiparticle to the satellites. The Dyson's equation governs the evolution of electron Green's function by repeated application of this self energy.
\begin{equation}
  G(n,\omega) = G_o(n,\omega) + G_o(n,\omega)\Sigma(\omega) G(n,\omega)
  \label{Dyson equation}
\end{equation}
%%%%%%%%%%%%%%%%%%%%%%%%%%%%%%%%%%%%%%%%%
\section{III. GW, Cumulant Expansion and Their Drawbacks}
The $GW$ approximation used to compute the quasiparticle properties are non self consistent and have abrupt truncation of Dyson's equation for computational efficiency unlike the fully self-consistent original $GW\Gamma$ formalism \cite{hedin_new_1965}. Although $GW$ based methods give reasonably good description of quasiparticle properties at weak coupling, the plasmon satellites are averaged and misplaced at some incorrect average energy \cite{langreth_singularities_1970}. At strong coupling, due to the lack of self-consistency, even the quasiparticle properties can be incorrect.

For a single (or isolated) band of electrons in a dispersionless plasmon bath \cite{langreth_singularities_1970}, an exact solution of the following form exists.
\begin{equation}
G(k,t) = G_o(k,t) e^{C(k,t)}
\label{cumulant}
\end{equation} 
The cumulant $C(k,t)$ is calculated by comparing equation \eqref{cumulant}'s taylor-expansion with the temporal Dyson's equation with $G$ and $\Sigma$ obtained from GW \cite{aryasetiawan_multiple_1996}. The satellites manifest as a Poisson series of peaks plasma frequency apart in the spectral function due to the exponential form of the cumulant ansatz \eqref{cumulant}. In real systems, although not all assumptions of above model hold true, an approximate cumulant correction can be found using the same recipe as above on a GW self energy. Recently, interest in the cumulant approximation has re-surged \cite{kas_cumulant_2014,gumhalter_combined_2016,caruso_band_2015,lischner_physical_2013} enabled by increases in computational ability to perform $GW$ and inspired by experiments (.e.g. \cite{guzzo_multiple_2014}) on complex systems.

The cumulant has the considerable merit of giving near-exact spectra for weak electron-boson coupling- $\Delta\gg\omega_o$ and/or $g\ll1$. However, at strong coupling and presence of multiple electronic levels, the bosons significantly affect the quasi-particle properties in ways not reflected in the cumulant approximation. The cumulant is also not systematically improvable by design and lacks proper accounting of inter-band scattering owing to the absence of self-consistency. 

\section{IV.Theoretical Framework:}
%\subsection{ Power Series Ansatz and its Properties %\protect}
\textbf{The Power Series Ansatz: }Rather than assuming an exponential correction, we assume a power series correction $\mathcal{P}(n,t)$ in the powers of  $g^2$ to the $n^{th}$ orbital's electron bare Green's function $G_o(n,t)$ due to interaction with bosons for time duration `t'. By construction, the interacting system smoothly maps to the non interacting system as $g^2$ goes to zero. 
\begin{equation}
\begin{aligned}
    G(n,t) = G_o(n,t) \mathcal{P}(n,t) = G_o(n,t)\sum_{k=0}^\infty g^{2k} C_k(n,t)
\end{aligned}
\label{eq:Corrected_Greens}
\end{equation}
Here $C_0 = 1$ and all other $C_k$ are distinct correction functions of different orders that are 0 when $t<0$. This makes physical sense because in retarded time framework- the particle doesn't exist for $t<0$. This, just like cumulant, is still a diagonal approximation to the Green's function matrix because, by construction, only those corrections in which a particle eventually returns back to its initial state $n$ are accounted for. 

\textbf{Temporal contraction relation: } For a given orbital n and time $t_i<t_o<t_f$, both $G$ and $G_o$ and hence by inheritance $\mathcal{P}$ have the following temporal contraction property due to the boundary value dependence on time. 
\begin{equation}
    f(n,t_f-t_i) = f(n,t_f-t_o)f(n,t_o-t_i)
    \label{contraction}
\end{equation}
This property doesn't apply between these function for different orbitals. In calculations, this seemingly trivial property of $\mathcal{P}(n,t)$ is absolutely essential to account for bosonic crossing diagrams. 

%%%%%%%%%%%%%%%%%%%%%%%%%%%%%%%%%%%%%%%%%%%%%%%%%%%%%%%%%%%%%%%%%%%%%%%%
%\subsection{Assumption on Electron Self Energy}
\textbf{Assumption on Electron Self Energy: }To properly construct the electron self energy, rather than replacing $G$ by $G_o$ inside the self energy as in $GW$ or cumulant expansions, we replace it by power series ansatz in order to re-introduce self-consistency.
%%%%%%%%%%%%%%%%%%%%%%%%%
%%%%%%%%%%%%%%%%%%%%%%%%%
\begin{equation}
\begin{aligned}
    -i\Sigma(t) &= g^2 \sum_{n=\pm}\,\sum_{N= \pm} \mathcal{D}(N,t)G_o(n, t)\mathcal{P}(n,t)\\
    &=g^2 \sum_{n=\pm}-i\Sigma_o(n,t)\mathcal{P}(n,t)
\end{aligned}
\label{Power corrected self energy}
\end{equation}
Here, the $n^{th}$ orbital's self-energy $\Sigma_o(n,t)$ computed by using bare Green's function $G_o$. The introduction of power series in $\Sigma$ through $G$ now produces corrections due to the particle's eventual return to the initial state after scattering through other possible states. Including these cyclic scattering contributions in the Green's function matrix's diagonal makes the diagonal exact.

%%%%%%%%%%%%%%%%%%%%%%%%%%%%%%%%%%%%%%%%%%%%%%%%%%%%%%%%%%%%%%%%%%%%%%%%%%%%%%%%%
%\subsection{Correction Scheme}
\textbf{Correction Scheme: }We take the temporal Dyson's equation for $m^{th}$ band and replace $G$ and $\Sigma$ by their power series corrected versions from \eqref{eq:Corrected_Greens}, \eqref{Power corrected self energy}. We then use the temporal limits enforced by the RT bare Green's function \eqref{Bare electron and boson greens and  greens} and simplify the equation using the temporal contraction property from equation \eqref{contraction}. 
 \begin{figure}
    \centering
    \includegraphics[height = 9 cm,width = 0.48\textwidth]{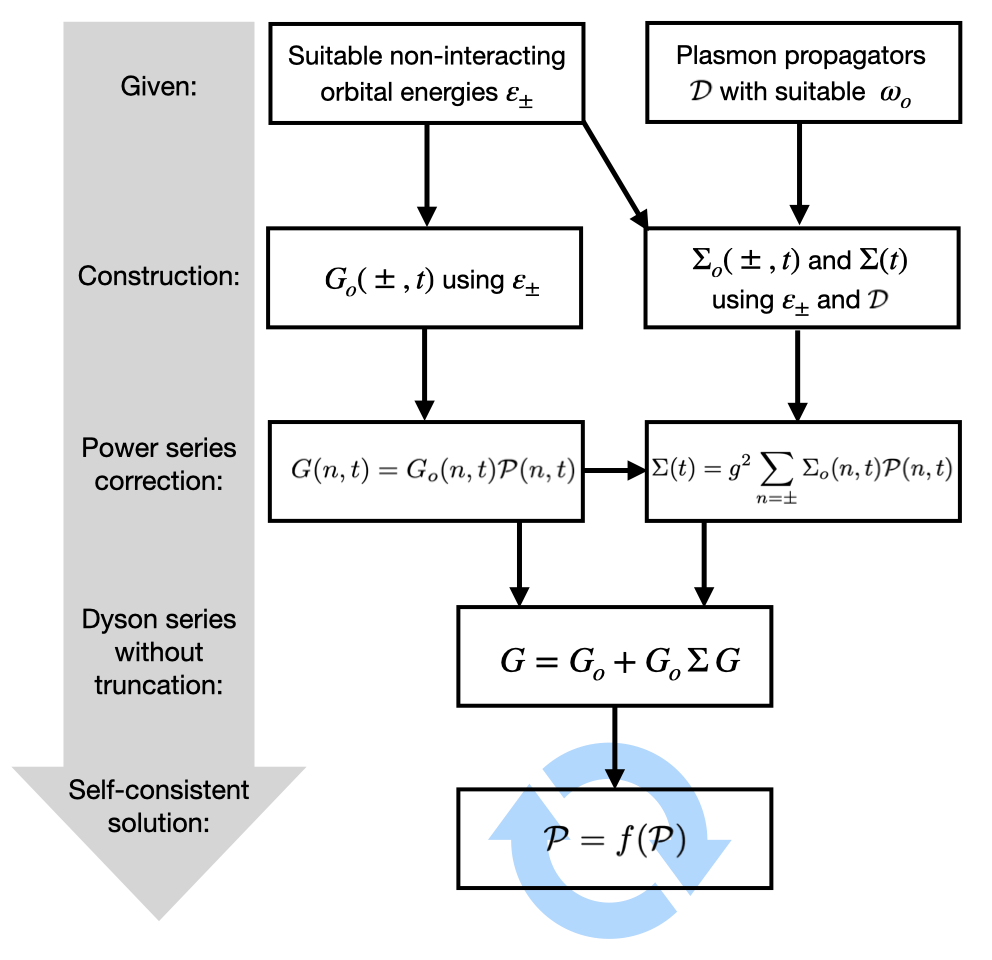}
    \caption{Flow chart of the Correction Scheme}
    \label{Flow Chart}
\end{figure}
\begin{equation*}
\begin{split}
     G(m,t-t_0) &=  G_o(m,t-t_0) \quad +\\
     \!\!\iint d{t_1} &d{t_2} G_o(m,t-t_2)\Sigma(t_2-t_1)G(m,t_1-t_0)
\end{split}
\end{equation*}
Setting $t_0=0$ and $t_2-t_1 = \tau$, and simplifying, we get,
\begin{equation*}
\begin{split}
    \mathcal{P}(m,t) =  \quad 1 &\quad+\\
    (-ig^2)\sum_{n=\pm}\int\displaylimits_{0}^{t} & \! dt_2 \!\!\int \displaylimits_{0}^{t_2}\!\!d \tau  e^{i\varepsilon_{m}\tau}\Sigma_o(n,\tau)\mathcal{P}(n,\tau) \mathcal{P}(m,t_2-\tau)
\end{split}
\end{equation*}
There are two distinct terms in this equation. The self correction ($P_{SC}$) term occurs when the interaction is within same orbital ($n=m$) on the right side of this equation. Here, the contraction property \eqref{contraction} must be used between the power series pieces on the right. The inter-band scattering term ($P_{IC}$) occurs when different orbitals interact ($n\neq m$) and here the contraction property is no longer valid. 
\begin{equation}
    \begin{aligned}
    &\therefore \mathcal{P}(m,t) =  1 + P_{SC} + P_{IC}\quad \quad\text{where,}\\
    &P_{SC}=\! -ig^2\!\!\int\displaylimits_{0}^{t}\!\!dt_2 \!\int\displaylimits_{0}^{t_2}\!\!d{\tau}\, e^{i\varepsilon_{m}\tau}\Sigma_o(m,\tau)\mathcal{P}(m,t_2)\\
     &P_{IC}= -ig^2\!\!\int\displaylimits_{0}^{t}\!\!dt_2 \!\!\int\displaylimits_{0}^{t_2}\!\!d\tau\, e^{i\varepsilon_{m}\tau} \,\Sigma_o(n,\tau)\mathcal{P}(n,\tau)\mathcal{P}(m,t_2-\tau)\\
    \end{aligned}
    \label{Power_Corr_full}
\end{equation}
For numerical solution, we start with an initial guess of $\mathcal{P} = 1$ on the right and self consistently compute better values for $\mathcal{P}$ on the left until it converges. 
%%%%%%%%%%%%%%%%%%%%%%%%%%%%%%%%%%%%%%%%%%%%%%

\section{V. Derivation of Various Cumulant Schemes }
We validate our method by deriving the exact cumulant result for the core-hole problem with single orbital of bare energy $\varepsilon_o$ in a bath of dispersionless plasmons of frequency $\omega_o$ \cite{langreth_singularities_1970}. The Hamiltonian in this case is;
\begin{equation*}
    \label{Core hole}
    H = \varepsilon_o c^\dagger c + \omega_o b^\dagger b + g (b^\dagger + b)(c^\dagger c -1)
\end{equation*}
This is an idealization of an isolated electron energy level $\varepsilon_o$ deep under the Fermi level being probed using x-ray photo-emission \cite{offi_comparison_2007}. The energetic electron exiting the system leaves behind a hole and the electron cloud responds to this imbalance of Coulomb forces by undergoing quantized long range oscillations (plasmons) at multiples of $\omega_o$. The corrected self energy for this case is;
\begin{equation*}
        \Sigma(t) \,= g^2 \Sigma_o(t)\mathcal{P}(t)\, =\, g^2 \big[-ie^{-i(\varepsilon_o-\omega_o)t} \theta(t)\big] \mathcal{P}(t)
\end{equation*}
For a single energy level, there is no inter-band scattering correction in equation \eqref{Power_Corr_full}.
\begin{equation*}
    \begin{aligned}
        \mathcal{P}(t)=1+ \Big[-ig^2\!\!\int\displaylimits_{0}^{t} dt_2 \int\displaylimits_{0}^{t_2}\!\!d{\tau}\, e^{i\varepsilon_{o}\tau}\Sigma_o(\tau)\mathcal{P}(t_2)\Big]
    \end{aligned}
    \label{single-disp-boson}
\end{equation*}

Expanding power series on both sides and comparing terms of same order in $g^2$ across the equality, we generate the following higher order corrections.
\begin{equation*}
    C_1(t) = \Big[\frac{e^{i\omega_o t} - i\omega_o t -1}{\omega_o^2}\Big] \quad \text{and}\quad C_k (t) = \frac{C_1(t)^k}{k!}
\end{equation*}
Summing all of these corrections gives us the exact result for the core hole problem.
\begin{equation}
G(t) = G_o(t)\mathcal{P}(t) = G_o(t)e^{g^2 C_1(t)}
\label{core-hole-cumulant}
\end{equation}

 The time-ordered cumulant expression in \cite{aryasetiawan_multiple_1996,gumhalter_combined_2016, caruso_band_2015,lischner_physical_2013} was derived assuming that the $n^{th}$ orbital's cumulant $C(n,t)$ depends only on the $n^{th}$ orbital's self energy $\Sigma(n,t)$ - thereby neglecting boson mediated inter-band scattering effects. In power series language, this translates as neglecting the effect of $H_-$ by setting $P_{IC}$ to be 0. In Holstein model, this means that the band gap $\Delta \gg \omega_o$ and each orbital essentially is an independent core-hole problem with corrections governed by $P_{SC}$ alone.

In the other limit - $\Delta \ll \omega_o$, the satellites are so far away that they don't modify the quasiparticle appreciably. Hence, both $P_{SC}$  and $P_{IC}$ are small and scale roughly equally. So they can be approximated as being independent of the orbital index in \eqref{Power_Corr_full}. This orbital independence lets us use the temporal contraction \eqref{contraction} for $P_{IC}$ regardless of orbital identity thereby giving RT cumulant correction \cite{kas_cumulant_2014,guzzo_multiple_2014}. The details of both derivations can be found in the supplement to this paper.
%%%%%%%%%%%%%%%%%%%%%%%%%%%%%%%%%%%%%%%%%%%%%%
\section{VI. Comparison with Exact Diagonalization Result}
\begin{figure}[htp]
    \includegraphics[height = 11.6 cm,width = 0.47\textwidth]{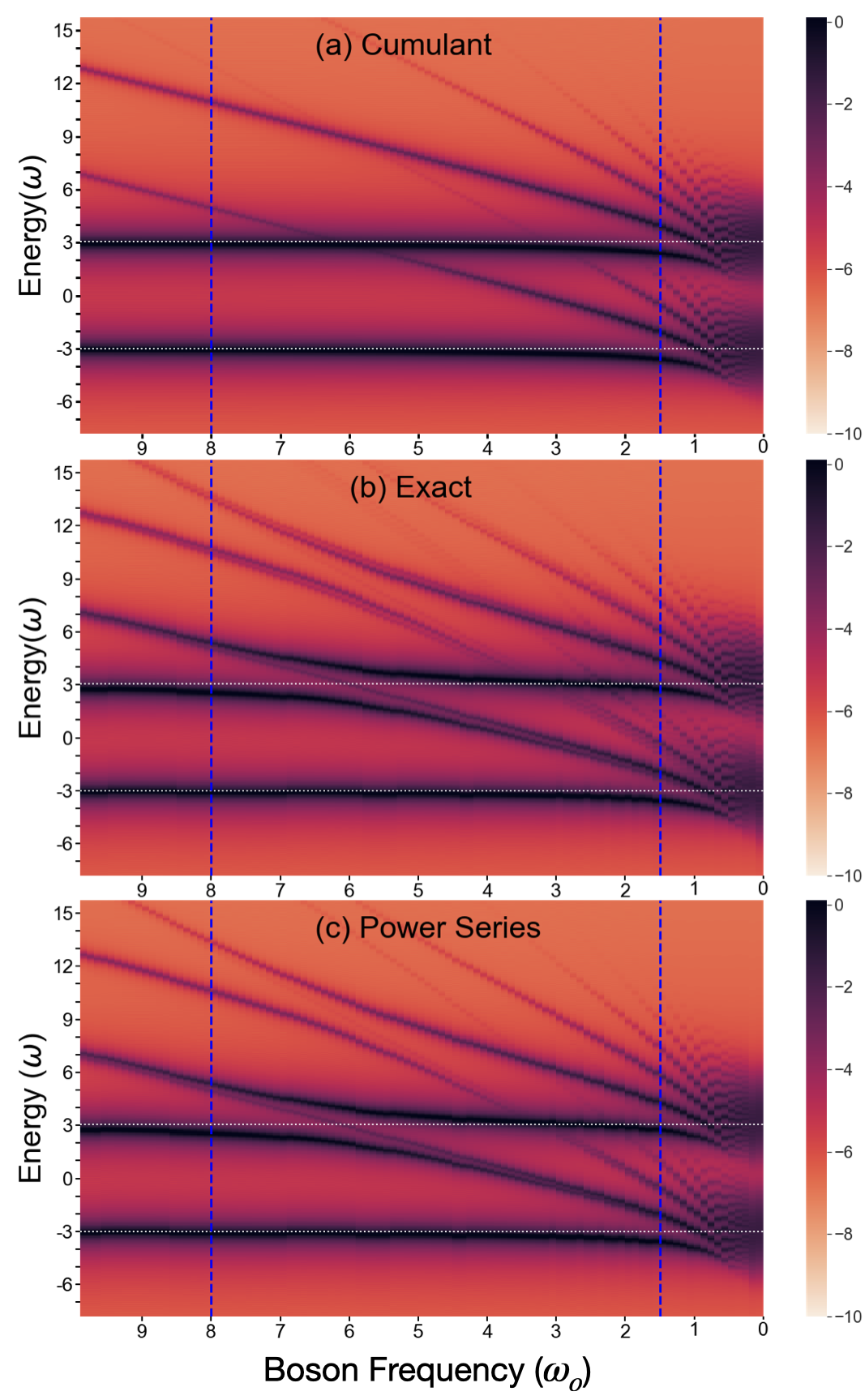}
    \caption{Natural log of spectral function from a) RT cumulant, b) exact diagonalization, and c) power series for $\varepsilon_\pm = \mp3$ (horizontal white dotted lines) and $\omega_o$ between 10 and 0.1. The blue vertical lines separate the three distinct regions.}
    \label{three graphs}
\end{figure}
We now numerically compute and compare the spectral functions obtained from power series, the exact diagonalization ($N\geq40$ boson basis) and RT cumulant for problem \eqref{Holstein} with $\varepsilon_\mp = \pm 3$, $\omega_o$ from 10 to 0.1, spectral broadening of 0.1, and a strong coupling parameter of $g=1$ in figure \ref{three graphs}. Depending on the magnitude of $\omega_o$ with respect to $\Delta$, figure \ref{three graphs} separates into three distinct regions roughly demarcated by the dashed blue lines.

 The first region is the weak coupling regime of $\omega_o\! \gg\! \Delta$ - here $\omega_o\!>\!8$. Here, both $(\pm)$ plasmon satellites are far away from the quasiparticle and therefore their effect on the quasiparticle energy and weight is negligible. This is most prominently seen from the negligible change quasiparticle energy from the non-interacting energies $\varepsilon_\pm$. Here, the retarded cumulant adequately captures all the exact spectral features correctly.
\begin{figure}[htp]
    \centering
    \includegraphics[height = 4.5 cm, width = 0.45\textwidth]{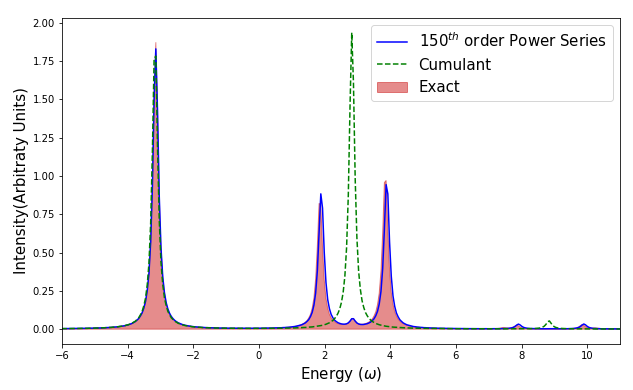}
    \caption{Spectral function with $g=1$, $\varepsilon_\pm = \mp3$, and $\omega_o =6$ from the three methods. Power series, unlike cumulant, captures the anti-bonding orbital splitting.}
    \label{omega_equals_delta}
\end{figure}

The second region has $\omega_o \approx \Delta$ - here $8>\omega_o>1.5$. A huge shift of spectral weight occurs from bonding to the anti-bonding orbital effectively splitting the anti-bonding orbital into two (between $\omega_o$ of 4 and 7). The shake-off replicas of this split level also come in pairs as seen in the exact spectra in figure \ref{omega_equals_delta}. These are captured exactly by the power series but not by cumulant because it lacks proper accounting of inter-band interaction.

The third region is when $\omega_o \ll \Delta$ - here $\omega_o < 1.5$. Here the bosonic events are extremely localized around the non-interacting energy and (+) bosons dominate the process. Therefore, inter-band interaction is vanishingly small and the solution is dominated by self-correction i.e core-hole like cumulant. We observe this in all three spectral functions although both exact and power series solution become computationally expensive -  the former due to large boson number necessary and the latter due to small time-step and large convergence order.

%%%%%%%%%%%%%%%%%%%%%%%%%%%%%%%%%%%%%%%%%%%%%%
\section{VII. Conclusion}
In this work, we derived a general power series based method which mitigates all the problems of cumulant-based methods, is practical to implement and reproduces the exact result in a finite basis for this problem within the spectral broadening used. We also identified three important regimes of this problem and elucidated where cumulant works, why cumulant works, and when it fails. We hope to extend this work to real multi-electron systems with strong plasmon resonances. 
 
%\nocite{*}
\bibliography{Power_Series_5}% Produces the bibliography via BibTeX.

%apsrev4-2.bst 2019-01-14 (MD) hand-edited version of apsrev4-1.bst
%Control: key (0)
%Control: author (72) initials jnrlst
%Control: editor formatted (1) identically to author
%Control: production of article title (-1) disabled
%Control: page (0) single
%Control: year (1) truncated
%Control: production of eprint (0) enabled
\begin{thebibliography}{31}%
\makeatletter
\providecommand \@ifxundefined [1]{%
 \@ifx{#1\undefined}
}%
\providecommand \@ifnum [1]{%
 \ifnum #1\expandafter \@firstoftwo
 \else \expandafter \@secondoftwo
 \fi
}%
\providecommand \@ifx [1]{%
 \ifx #1\expandafter \@firstoftwo
 \else \expandafter \@secondoftwo
 \fi
}%
\providecommand \natexlab [1]{#1}%
\providecommand \enquote  [1]{``#1''}%
\providecommand \bibnamefont  [1]{#1}%
\providecommand \bibfnamefont [1]{#1}%
\providecommand \citenamefont [1]{#1}%
\providecommand \href@noop [0]{\@secondoftwo}%
\providecommand \href [0]{\begingroup \@sanitize@url \@href}%
\providecommand \@href[1]{\@@startlink{#1}\@@href}%
\providecommand \@@href[1]{\endgroup#1\@@endlink}%
\providecommand \@sanitize@url [0]{\catcode `\\12\catcode `\$12\catcode
  `\&12\catcode `\#12\catcode `\^12\catcode `\_12\catcode `\%12\relax}%
\providecommand \@@startlink[1]{}%
\providecommand \@@endlink[0]{}%
\providecommand \url  [0]{\begingroup\@sanitize@url \@url }%
\providecommand \@url [1]{\endgroup\@href {#1}{\urlprefix }}%
\providecommand \urlprefix  [0]{URL }%
\providecommand \Eprint [0]{\href }%
\providecommand \doibase [0]{https://doi.org/}%
\providecommand \selectlanguage [0]{\@gobble}%
\providecommand \bibinfo  [0]{\@secondoftwo}%
\providecommand \bibfield  [0]{\@secondoftwo}%
\providecommand \translation [1]{[#1]}%
\providecommand \BibitemOpen [0]{}%
\providecommand \bibitemStop [0]{}%
\providecommand \bibitemNoStop [0]{.\EOS\space}%
\providecommand \EOS [0]{\spacefactor3000\relax}%
\providecommand \BibitemShut  [1]{\csname bibitem#1\endcsname}%
\let\auto@bib@innerbib\@empty
%</preamble>
\bibitem [{\citenamefont {Goulielmakis}\ \emph {et~al.}(2010)\citenamefont
  {Goulielmakis}, \citenamefont {Loh}, \citenamefont {Wirth}, \citenamefont
  {Santra}, \citenamefont {Rohringer}, \citenamefont {Yakovlev}, \citenamefont
  {Zherebtsov}, \citenamefont {Pfeifer}, \citenamefont {Azzeer}, \citenamefont
  {Kling}, \citenamefont {Leone},\ and\ \citenamefont
  {Krausz}}]{goulielmakis_real-time_2010}%
  \BibitemOpen
  \bibfield  {author} {\bibinfo {author} {\bibfnamefont {E.}~\bibnamefont
  {Goulielmakis}}, \bibinfo {author} {\bibfnamefont {Z.-H.}\ \bibnamefont
  {Loh}}, \bibinfo {author} {\bibfnamefont {A.}~\bibnamefont {Wirth}}, \bibinfo
  {author} {\bibfnamefont {R.}~\bibnamefont {Santra}}, \bibinfo {author}
  {\bibfnamefont {N.}~\bibnamefont {Rohringer}}, \bibinfo {author}
  {\bibfnamefont {V.~S.}\ \bibnamefont {Yakovlev}}, \bibinfo {author}
  {\bibfnamefont {S.}~\bibnamefont {Zherebtsov}}, \bibinfo {author}
  {\bibfnamefont {T.}~\bibnamefont {Pfeifer}}, \bibinfo {author} {\bibfnamefont
  {A.~M.}\ \bibnamefont {Azzeer}}, \bibinfo {author} {\bibfnamefont {M.~F.}\
  \bibnamefont {Kling}}, \bibinfo {author} {\bibfnamefont {S.~R.}\ \bibnamefont
  {Leone}},\ and\ \bibinfo {author} {\bibfnamefont {F.}~\bibnamefont
  {Krausz}},\ }\bibfield  {journal} {\bibinfo  {journal} {Nature}\ }\textbf
  {\bibinfo {volume} {466}},\ \href {https://doi.org/10.1038/nature09212}
  {10.1038/nature09212} (\bibinfo {year} {2010})\BibitemShut {NoStop}%
\bibitem [{\citenamefont {Lemell}\ \emph {et~al.}(2015)\citenamefont {Lemell},
  \citenamefont {Neppl}, \citenamefont {Wachter}, \citenamefont {Tőkési},
  \citenamefont {Ernstorfer}, \citenamefont {Feulner}, \citenamefont
  {Kienberger},\ and\ \citenamefont {Burgdörfer}}]{lemell_real-time_2015}%
  \BibitemOpen
  \bibfield  {author} {\bibinfo {author} {\bibfnamefont {C.}~\bibnamefont
  {Lemell}}, \bibinfo {author} {\bibfnamefont {S.}~\bibnamefont {Neppl}},
  \bibinfo {author} {\bibfnamefont {G.}~\bibnamefont {Wachter}}, \bibinfo
  {author} {\bibfnamefont {K.}~\bibnamefont {Tőkési}}, \bibinfo {author}
  {\bibfnamefont {R.}~\bibnamefont {Ernstorfer}}, \bibinfo {author}
  {\bibfnamefont {P.}~\bibnamefont {Feulner}}, \bibinfo {author} {\bibfnamefont
  {R.}~\bibnamefont {Kienberger}},\ and\ \bibinfo {author} {\bibfnamefont
  {J.}~\bibnamefont {Burgdörfer}},\ }\bibfield  {journal} {\bibinfo  {journal}
  {Physical Review B}\ }\textbf {\bibinfo {volume} {91}},\ \href
  {https://doi.org/10.1103/PhysRevB.91.241101} {10.1103/PhysRevB.91.241101}
  (\bibinfo {year} {2015})\BibitemShut {NoStop}%
\bibitem [{\citenamefont {Neppl}\ \emph {et~al.}(2015)\citenamefont {Neppl},
  \citenamefont {Ernstorfer}, \citenamefont {Cavalieri}, \citenamefont
  {Lemell}, \citenamefont {Wachter}, \citenamefont {Magerl}, \citenamefont
  {Bothschafter}, \citenamefont {Jobst}, \citenamefont {Hofstetter},
  \citenamefont {Kleineberg}, \citenamefont {Barth}, \citenamefont {Menzel},
  \citenamefont {Burgdörfer}, \citenamefont {Feulner}, \citenamefont
  {Krausz},\ and\ \citenamefont {Kienberger}}]{neppl_direct_2015}%
  \BibitemOpen
  \bibfield  {author} {\bibinfo {author} {\bibfnamefont {S.}~\bibnamefont
  {Neppl}}, \bibinfo {author} {\bibfnamefont {R.}~\bibnamefont {Ernstorfer}},
  \bibinfo {author} {\bibfnamefont {A.~L.}\ \bibnamefont {Cavalieri}}, \bibinfo
  {author} {\bibfnamefont {C.}~\bibnamefont {Lemell}}, \bibinfo {author}
  {\bibfnamefont {G.}~\bibnamefont {Wachter}}, \bibinfo {author} {\bibfnamefont
  {E.}~\bibnamefont {Magerl}}, \bibinfo {author} {\bibfnamefont {E.~M.}\
  \bibnamefont {Bothschafter}}, \bibinfo {author} {\bibfnamefont
  {M.}~\bibnamefont {Jobst}}, \bibinfo {author} {\bibfnamefont
  {M.}~\bibnamefont {Hofstetter}}, \bibinfo {author} {\bibfnamefont
  {U.}~\bibnamefont {Kleineberg}}, \bibinfo {author} {\bibfnamefont {J.~V.}\
  \bibnamefont {Barth}}, \bibinfo {author} {\bibfnamefont {D.}~\bibnamefont
  {Menzel}}, \bibinfo {author} {\bibfnamefont {J.}~\bibnamefont {Burgdörfer}},
  \bibinfo {author} {\bibfnamefont {P.}~\bibnamefont {Feulner}}, \bibinfo
  {author} {\bibfnamefont {F.}~\bibnamefont {Krausz}},\ and\ \bibinfo {author}
  {\bibfnamefont {R.}~\bibnamefont {Kienberger}},\ }\bibfield  {journal}
  {\bibinfo  {journal} {Nature}\ }\textbf {\bibinfo {volume} {517}},\ \href
  {https://doi.org/10.1038/nature14094} {10.1038/nature14094} (\bibinfo {year}
  {2015})\BibitemShut {NoStop}%
\bibitem [{\citenamefont {Wang}\ \emph
  {et~al.}(2016{\natexlab{a}})\citenamefont {Wang}, \citenamefont
  {McKeown~Walker}, \citenamefont {Tamai}, \citenamefont {Wang}, \citenamefont
  {Ristic}, \citenamefont {Bruno}, \citenamefont {de~la Torre}, \citenamefont
  {Riccò}, \citenamefont {Plumb}, \citenamefont {Shi}, \citenamefont
  {Hlawenka}, \citenamefont {Sánchez-Barriga}, \citenamefont {Varykhalov},
  \citenamefont {Kim}, \citenamefont {Hoesch}, \citenamefont {King},
  \citenamefont {Meevasana}, \citenamefont {Diebold}, \citenamefont {Mesot},
  \citenamefont {Moritz}, \citenamefont {Devereaux}, \citenamefont {Radovic},\
  and\ \citenamefont {Baumberger}}]{wang_tailoring_2016}%
  \BibitemOpen
  \bibfield  {author} {\bibinfo {author} {\bibfnamefont {Z.}~\bibnamefont
  {Wang}}, \bibinfo {author} {\bibfnamefont {S.}~\bibnamefont
  {McKeown~Walker}}, \bibinfo {author} {\bibfnamefont {A.}~\bibnamefont
  {Tamai}}, \bibinfo {author} {\bibfnamefont {Y.}~\bibnamefont {Wang}},
  \bibinfo {author} {\bibfnamefont {Z.}~\bibnamefont {Ristic}}, \bibinfo
  {author} {\bibfnamefont {F.~Y.}\ \bibnamefont {Bruno}}, \bibinfo {author}
  {\bibfnamefont {A.}~\bibnamefont {de~la Torre}}, \bibinfo {author}
  {\bibfnamefont {S.}~\bibnamefont {Riccò}}, \bibinfo {author} {\bibfnamefont
  {N.~C.}\ \bibnamefont {Plumb}}, \bibinfo {author} {\bibfnamefont
  {M.}~\bibnamefont {Shi}}, \bibinfo {author} {\bibfnamefont {P.}~\bibnamefont
  {Hlawenka}}, \bibinfo {author} {\bibfnamefont {J.}~\bibnamefont
  {Sánchez-Barriga}}, \bibinfo {author} {\bibfnamefont {A.}~\bibnamefont
  {Varykhalov}}, \bibinfo {author} {\bibfnamefont {T.~K.}\ \bibnamefont {Kim}},
  \bibinfo {author} {\bibfnamefont {M.}~\bibnamefont {Hoesch}}, \bibinfo
  {author} {\bibfnamefont {P.~D.~C.}\ \bibnamefont {King}}, \bibinfo {author}
  {\bibfnamefont {W.}~\bibnamefont {Meevasana}}, \bibinfo {author}
  {\bibfnamefont {U.}~\bibnamefont {Diebold}}, \bibinfo {author} {\bibfnamefont
  {J.}~\bibnamefont {Mesot}}, \bibinfo {author} {\bibfnamefont
  {B.}~\bibnamefont {Moritz}}, \bibinfo {author} {\bibfnamefont {T.~P.}\
  \bibnamefont {Devereaux}}, \bibinfo {author} {\bibfnamefont {M.}~\bibnamefont
  {Radovic}},\ and\ \bibinfo {author} {\bibfnamefont {F.}~\bibnamefont
  {Baumberger}},\ }\bibfield  {journal} {\bibinfo  {journal} {Nature
  Materials}\ }\textbf {\bibinfo {volume} {15}},\ \href
  {https://doi.org/10.1038/nmat4623} {10.1038/nmat4623} (\bibinfo {year}
  {2016}{\natexlab{a}})\BibitemShut {NoStop}%
\bibitem [{\citenamefont {Kang}\ \emph {et~al.}(2018)\citenamefont {Kang},
  \citenamefont {Jung}, \citenamefont {Shin}, \citenamefont {Sohn},
  \citenamefont {Ryu}, \citenamefont {Kim}, \citenamefont {Hoesch},\ and\
  \citenamefont {Kim}}]{kang_holstein_2018}%
  \BibitemOpen
  \bibfield  {author} {\bibinfo {author} {\bibfnamefont {M.}~\bibnamefont
  {Kang}}, \bibinfo {author} {\bibfnamefont {S.~W.}\ \bibnamefont {Jung}},
  \bibinfo {author} {\bibfnamefont {W.~J.}\ \bibnamefont {Shin}}, \bibinfo
  {author} {\bibfnamefont {Y.}~\bibnamefont {Sohn}}, \bibinfo {author}
  {\bibfnamefont {S.~H.}\ \bibnamefont {Ryu}}, \bibinfo {author} {\bibfnamefont
  {T.~K.}\ \bibnamefont {Kim}}, \bibinfo {author} {\bibfnamefont
  {M.}~\bibnamefont {Hoesch}},\ and\ \bibinfo {author} {\bibfnamefont {K.~S.}\
  \bibnamefont {Kim}},\ }\bibfield  {journal} {\bibinfo  {journal} {Nature
  Materials}\ }\textbf {\bibinfo {volume} {17}},\ \href
  {https://doi.org/10.1038/s41563-018-0092-7} {10.1038/s41563-018-0092-7}
  (\bibinfo {year} {2018})\BibitemShut {NoStop}%
\bibitem [{\citenamefont {van Mechelen}\ \emph {et~al.}(2008)\citenamefont {van
  Mechelen}, \citenamefont {van~der Marel}, \citenamefont {Grimaldi},
  \citenamefont {Kuzmenko}, \citenamefont {Armitage}, \citenamefont {Reyren},
  \citenamefont {Hagemann},\ and\ \citenamefont
  {Mazin}}]{van_mechelen_electron-phonon_2008}%
  \BibitemOpen
  \bibfield  {author} {\bibinfo {author} {\bibfnamefont {J.~L.~M.}\
  \bibnamefont {van Mechelen}}, \bibinfo {author} {\bibfnamefont
  {D.}~\bibnamefont {van~der Marel}}, \bibinfo {author} {\bibfnamefont
  {C.}~\bibnamefont {Grimaldi}}, \bibinfo {author} {\bibfnamefont {A.~B.}\
  \bibnamefont {Kuzmenko}}, \bibinfo {author} {\bibfnamefont {N.~P.}\
  \bibnamefont {Armitage}}, \bibinfo {author} {\bibfnamefont {N.}~\bibnamefont
  {Reyren}}, \bibinfo {author} {\bibfnamefont {H.}~\bibnamefont {Hagemann}},\
  and\ \bibinfo {author} {\bibfnamefont {I.~I.}\ \bibnamefont {Mazin}},\
  }\bibfield  {journal} {\bibinfo  {journal} {Physical Review Letters}\
  }\textbf {\bibinfo {volume} {100}},\ \href
  {https://doi.org/10.1103/PhysRevLett.100.226403}
  {10.1103/PhysRevLett.100.226403} (\bibinfo {year} {2008})\BibitemShut
  {NoStop}%
\bibitem [{\citenamefont {Devreese}(2010)}]{devreese_many-body_2010}%
  \BibitemOpen
  \bibfield  {author} {\bibinfo {author} {\bibfnamefont {J.~T.}\ \bibnamefont
  {Devreese}},\ }\bibfield  {journal} {\bibinfo  {journal} {Physical Review B}\
  }\textbf {\bibinfo {volume} {81}},\ \href
  {https://doi.org/10.1103/PhysRevB.81.125119} {10.1103/PhysRevB.81.125119}
  (\bibinfo {year} {2010})\BibitemShut {NoStop}%
\bibitem [{\citenamefont {Wang}\ \emph
  {et~al.}(2016{\natexlab{b}})\citenamefont {Wang}, \citenamefont
  {McKeown~Walker}, \citenamefont {Tamai}, \citenamefont {Wang}, \citenamefont
  {Ristic}, \citenamefont {Bruno}, \citenamefont {de~la Torre}, \citenamefont
  {Ricc{\`o}}, \citenamefont {Plumb}, \citenamefont {Shi}, \citenamefont
  {Hlawenka}, \citenamefont {S{\'a}nchez-Barriga}, \citenamefont {Varykhalov},
  \citenamefont {Kim}, \citenamefont {Hoesch}, \citenamefont {King},
  \citenamefont {Meevasana}, \citenamefont {Diebold}, \citenamefont {Mesot},
  \citenamefont {Moritz}, \citenamefont {Devereaux}, \citenamefont {Radovic},\
  and\ \citenamefont {Baumberger}}]{wang2016}%
  \BibitemOpen
  \bibfield  {author} {\bibinfo {author} {\bibfnamefont {Z.}~\bibnamefont
  {Wang}}, \bibinfo {author} {\bibfnamefont {S.}~\bibnamefont
  {McKeown~Walker}}, \bibinfo {author} {\bibfnamefont {A.}~\bibnamefont
  {Tamai}}, \bibinfo {author} {\bibfnamefont {Y.}~\bibnamefont {Wang}},
  \bibinfo {author} {\bibfnamefont {Z.}~\bibnamefont {Ristic}}, \bibinfo
  {author} {\bibfnamefont {F.~Y.}\ \bibnamefont {Bruno}}, \bibinfo {author}
  {\bibfnamefont {A.}~\bibnamefont {de~la Torre}}, \bibinfo {author}
  {\bibfnamefont {S.}~\bibnamefont {Ricc{\`o}}}, \bibinfo {author}
  {\bibfnamefont {N.~C.}\ \bibnamefont {Plumb}}, \bibinfo {author}
  {\bibfnamefont {M.}~\bibnamefont {Shi}}, \bibinfo {author} {\bibfnamefont
  {P.}~\bibnamefont {Hlawenka}}, \bibinfo {author} {\bibfnamefont
  {J.}~\bibnamefont {S{\'a}nchez-Barriga}}, \bibinfo {author} {\bibfnamefont
  {A.}~\bibnamefont {Varykhalov}}, \bibinfo {author} {\bibfnamefont {T.~K.}\
  \bibnamefont {Kim}}, \bibinfo {author} {\bibfnamefont {M.}~\bibnamefont
  {Hoesch}}, \bibinfo {author} {\bibfnamefont {P.~D.~C.}\ \bibnamefont {King}},
  \bibinfo {author} {\bibfnamefont {W.}~\bibnamefont {Meevasana}}, \bibinfo
  {author} {\bibfnamefont {U.}~\bibnamefont {Diebold}}, \bibinfo {author}
  {\bibfnamefont {J.}~\bibnamefont {Mesot}}, \bibinfo {author} {\bibfnamefont
  {B.}~\bibnamefont {Moritz}}, \bibinfo {author} {\bibfnamefont {T.~P.}\
  \bibnamefont {Devereaux}}, \bibinfo {author} {\bibfnamefont {M.}~\bibnamefont
  {Radovic}},\ and\ \bibinfo {author} {\bibfnamefont {F.}~\bibnamefont
  {Baumberger}},\ }\href {https://doi.org/10.1038/nmat4623} {\bibfield
  {journal} {\bibinfo  {journal} {Nature Materials}\ }\textbf {\bibinfo
  {volume} {15}},\ \bibinfo {pages} {835} (\bibinfo {year}
  {2016}{\natexlab{b}})}\BibitemShut {NoStop}%
\bibitem [{\citenamefont {Swartz}\ \emph {et~al.}(2018)\citenamefont {Swartz},
  \citenamefont {Inoue}, \citenamefont {Merz}, \citenamefont {Hikita},
  \citenamefont {Raghu}, \citenamefont {Devereaux}, \citenamefont {Johnston},\
  and\ \citenamefont {Hwang}}]{swartz2018}%
  \BibitemOpen
  \bibfield  {author} {\bibinfo {author} {\bibfnamefont {A.~G.}\ \bibnamefont
  {Swartz}}, \bibinfo {author} {\bibfnamefont {H.}~\bibnamefont {Inoue}},
  \bibinfo {author} {\bibfnamefont {T.~A.}\ \bibnamefont {Merz}}, \bibinfo
  {author} {\bibfnamefont {Y.}~\bibnamefont {Hikita}}, \bibinfo {author}
  {\bibfnamefont {S.}~\bibnamefont {Raghu}}, \bibinfo {author} {\bibfnamefont
  {T.~P.}\ \bibnamefont {Devereaux}}, \bibinfo {author} {\bibfnamefont
  {S.}~\bibnamefont {Johnston}},\ and\ \bibinfo {author} {\bibfnamefont
  {H.~Y.}\ \bibnamefont {Hwang}},\ }\href
  {https://doi.org/10.1073/pnas.1713916115} {\bibfield  {journal} {\bibinfo
  {journal} {Proceedings of the National Academy of Sciences}\ }\textbf
  {\bibinfo {volume} {115}},\ \bibinfo {pages} {1475} (\bibinfo {year}
  {2018})},\ \Eprint
  {https://arxiv.org/abs/https://www.pnas.org/content/115/7/1475.full.pdf}
  {https://www.pnas.org/content/115/7/1475.full.pdf} \BibitemShut {NoStop}%
\bibitem [{\citenamefont {Edelman}\ and\ \citenamefont
  {Littlewood}(2021)}]{edelman2021}%
  \BibitemOpen
  \bibfield  {author} {\bibinfo {author} {\bibfnamefont {A.}~\bibnamefont
  {Edelman}}\ and\ \bibinfo {author} {\bibfnamefont {P.~B.}\ \bibnamefont
  {Littlewood}},\ }\href@noop {} {\bibinfo {title} {Normal state correlates of
  plasmon-polaron superconductivity in strontium titanate}} (\bibinfo {year}
  {2021}),\ \Eprint {https://arxiv.org/abs/2111.03138} {arXiv:2111.03138
  [cond-mat.supr-con]} \BibitemShut {NoStop}%
\bibitem [{\citenamefont {Rösch}\ \emph {et~al.}(2005)\citenamefont {Rösch},
  \citenamefont {Gunnarsson}, \citenamefont {Zhou}, \citenamefont {Yoshida},
  \citenamefont {Sasagawa}, \citenamefont {Fujimori}, \citenamefont {Hussain},
  \citenamefont {Shen},\ and\ \citenamefont {Uchida}}]{rosch_polaronic_2005}%
  \BibitemOpen
  \bibfield  {author} {\bibinfo {author} {\bibfnamefont {O.}~\bibnamefont
  {Rösch}}, \bibinfo {author} {\bibfnamefont {O.}~\bibnamefont {Gunnarsson}},
  \bibinfo {author} {\bibfnamefont {X.~J.}\ \bibnamefont {Zhou}}, \bibinfo
  {author} {\bibfnamefont {T.}~\bibnamefont {Yoshida}}, \bibinfo {author}
  {\bibfnamefont {T.}~\bibnamefont {Sasagawa}}, \bibinfo {author}
  {\bibfnamefont {A.}~\bibnamefont {Fujimori}}, \bibinfo {author}
  {\bibfnamefont {Z.}~\bibnamefont {Hussain}}, \bibinfo {author} {\bibfnamefont
  {Z.-X.}\ \bibnamefont {Shen}},\ and\ \bibinfo {author} {\bibfnamefont
  {S.}~\bibnamefont {Uchida}},\ }\bibfield  {journal} {\bibinfo  {journal}
  {Physical Review Letters}\ }\textbf {\bibinfo {volume} {95}},\ \href
  {https://doi.org/10.1103/PhysRevLett.95.227002}
  {10.1103/PhysRevLett.95.227002} (\bibinfo {year} {2005})\BibitemShut
  {NoStop}%
\bibitem [{\citenamefont {Damascelli}\ \emph {et~al.}(2003)\citenamefont
  {Damascelli}, \citenamefont {Hussain},\ and\ \citenamefont
  {Shen}}]{damascelli_angle-resolved_2003}%
  \BibitemOpen
  \bibfield  {author} {\bibinfo {author} {\bibfnamefont {A.}~\bibnamefont
  {Damascelli}}, \bibinfo {author} {\bibfnamefont {Z.}~\bibnamefont
  {Hussain}},\ and\ \bibinfo {author} {\bibfnamefont {Z.-X.}\ \bibnamefont
  {Shen}},\ }\bibfield  {journal} {\bibinfo  {journal} {Reviews of Modern
  Physics}\ }\textbf {\bibinfo {volume} {75}},\ \href
  {https://doi.org/10.1103/RevModPhys.75.473} {10.1103/RevModPhys.75.473}
  (\bibinfo {year} {2003})\BibitemShut {NoStop}%
\bibitem [{\citenamefont {Baldini}\ \emph {et~al.}(2020)\citenamefont
  {Baldini}, \citenamefont {Sentef}, \citenamefont {Acharya}, \citenamefont
  {Brumme}, \citenamefont {Sheveleva}, \citenamefont {Lyzwa}, \citenamefont
  {Pomjakushina}, \citenamefont {Bernhard}, \citenamefont {Schilfgaarde},
  \citenamefont {Carbone}, \citenamefont {Rubio},\ and\ \citenamefont
  {Weber}}]{baldini_electronphonon-driven_2020}%
  \BibitemOpen
  \bibfield  {author} {\bibinfo {author} {\bibfnamefont {E.}~\bibnamefont
  {Baldini}}, \bibinfo {author} {\bibfnamefont {M.~A.}\ \bibnamefont {Sentef}},
  \bibinfo {author} {\bibfnamefont {S.}~\bibnamefont {Acharya}}, \bibinfo
  {author} {\bibfnamefont {T.}~\bibnamefont {Brumme}}, \bibinfo {author}
  {\bibfnamefont {E.}~\bibnamefont {Sheveleva}}, \bibinfo {author}
  {\bibfnamefont {F.}~\bibnamefont {Lyzwa}}, \bibinfo {author} {\bibfnamefont
  {E.}~\bibnamefont {Pomjakushina}}, \bibinfo {author} {\bibfnamefont
  {C.}~\bibnamefont {Bernhard}}, \bibinfo {author} {\bibfnamefont {M.~v.}\
  \bibnamefont {Schilfgaarde}}, \bibinfo {author} {\bibfnamefont
  {F.}~\bibnamefont {Carbone}}, \bibinfo {author} {\bibfnamefont
  {A.}~\bibnamefont {Rubio}},\ and\ \bibinfo {author} {\bibfnamefont
  {C.}~\bibnamefont {Weber}},\ }\href {https://doi.org/10.1073/pnas.1919451117}
  {\bibfield  {journal} {\bibinfo  {journal} {Proceedings of the National
  Academy of Sciences}\ }\textbf {\bibinfo {volume} {117}},\ \bibinfo {pages}
  {6409} (\bibinfo {year} {2020})}\BibitemShut {NoStop}%
\bibitem [{\citenamefont {Yang}\ \emph {et~al.}(2015)\citenamefont {Yang},
  \citenamefont {Sobota}, \citenamefont {Leuenberger}, \citenamefont {Kemper},
  \citenamefont {Lee}, \citenamefont {Schmitt}, \citenamefont {Li},
  \citenamefont {Moore}, \citenamefont {Kirchmann},\ and\ \citenamefont
  {Shen}}]{yang2015}%
  \BibitemOpen
  \bibfield  {author} {\bibinfo {author} {\bibfnamefont {S.}~\bibnamefont
  {Yang}}, \bibinfo {author} {\bibfnamefont {J.~A.}\ \bibnamefont {Sobota}},
  \bibinfo {author} {\bibfnamefont {D.}~\bibnamefont {Leuenberger}}, \bibinfo
  {author} {\bibfnamefont {A.~F.}\ \bibnamefont {Kemper}}, \bibinfo {author}
  {\bibfnamefont {J.~J.}\ \bibnamefont {Lee}}, \bibinfo {author} {\bibfnamefont
  {F.~T.}\ \bibnamefont {Schmitt}}, \bibinfo {author} {\bibfnamefont
  {W.}~\bibnamefont {Li}}, \bibinfo {author} {\bibfnamefont {R.~G.}\
  \bibnamefont {Moore}}, \bibinfo {author} {\bibfnamefont {P.~S.}\ \bibnamefont
  {Kirchmann}},\ and\ \bibinfo {author} {\bibfnamefont {Z.-X.}\ \bibnamefont
  {Shen}},\ }\href {https://doi.org/10.1021/acs.nanolett.5b01274} {\bibfield
  {journal} {\bibinfo  {journal} {Nano Letters}\ }\textbf {\bibinfo {volume}
  {15}},\ \bibinfo {pages} {4150} (\bibinfo {year} {2015})},\ \bibinfo {note}
  {pMID: 26027951},\ \Eprint
  {https://arxiv.org/abs/https://doi.org/10.1021/acs.nanolett.5b01274}
  {https://doi.org/10.1021/acs.nanolett.5b01274} \BibitemShut {NoStop}%
\bibitem [{\citenamefont {Mohamed}\ \emph {et~al.}(2019)\citenamefont
  {Mohamed}, \citenamefont {M. May}, \citenamefont {Kanis}, \citenamefont
  {Brützam}, \citenamefont {Uecker}, \citenamefont {Krol}, \citenamefont
  {Janowitz},\ and\ \citenamefont {Mulazzi}}]{mohamed_electronic_2019}%
  \BibitemOpen
  \bibfield  {author} {\bibinfo {author} {\bibfnamefont {M.}~\bibnamefont
  {Mohamed}}, \bibinfo {author} {\bibfnamefont {M.}~\bibnamefont {M. May}},
  \bibinfo {author} {\bibfnamefont {M.}~\bibnamefont {Kanis}}, \bibinfo
  {author} {\bibfnamefont {M.}~\bibnamefont {Brützam}}, \bibinfo {author}
  {\bibfnamefont {R.}~\bibnamefont {Uecker}}, \bibinfo {author} {\bibfnamefont
  {R.~v.~d.}\ \bibnamefont {Krol}}, \bibinfo {author} {\bibfnamefont
  {C.}~\bibnamefont {Janowitz}},\ and\ \bibinfo {author} {\bibfnamefont
  {M.}~\bibnamefont {Mulazzi}},\ }\bibfield  {journal} {\bibinfo  {journal}
  {RSC Advances}\ }\textbf {\bibinfo {volume} {9}},\ \href
  {https://doi.org/10.1039/C9RA01092K} {10.1039/C9RA01092K} (\bibinfo {year}
  {2019})\BibitemShut {NoStop}%
\bibitem [{\citenamefont {Ma}\ \emph {et~al.}(2016)\citenamefont {Ma},
  \citenamefont {Dai}, \citenamefont {Yu},\ and\ \citenamefont
  {Huang}}]{ma_energy_2016}%
  \BibitemOpen
  \bibfield  {author} {\bibinfo {author} {\bibfnamefont {X.-C.}\ \bibnamefont
  {Ma}}, \bibinfo {author} {\bibfnamefont {Y.}~\bibnamefont {Dai}}, \bibinfo
  {author} {\bibfnamefont {L.}~\bibnamefont {Yu}},\ and\ \bibinfo {author}
  {\bibfnamefont {B.-B.}\ \bibnamefont {Huang}},\ }\bibfield  {journal}
  {\bibinfo  {journal} {Light: Science \& Applications}\ }\textbf {\bibinfo
  {volume} {5}},\ \href {https://doi.org/10.1038/lsa.2016.17}
  {10.1038/lsa.2016.17} (\bibinfo {year} {2016})\BibitemShut {NoStop}%
\bibitem [{\citenamefont {Hulea}\ \emph {et~al.}(2006)\citenamefont {Hulea},
  \citenamefont {Fratini}, \citenamefont {Xie}, \citenamefont {Mulder},
  \citenamefont {Iossad}, \citenamefont {Rastelli}, \citenamefont {Ciuchi},\
  and\ \citenamefont {Morpurgo}}]{hulea_tunable_2006}%
  \BibitemOpen
  \bibfield  {author} {\bibinfo {author} {\bibfnamefont {I.~N.}\ \bibnamefont
  {Hulea}}, \bibinfo {author} {\bibfnamefont {S.}~\bibnamefont {Fratini}},
  \bibinfo {author} {\bibfnamefont {H.}~\bibnamefont {Xie}}, \bibinfo {author}
  {\bibfnamefont {C.~L.}\ \bibnamefont {Mulder}}, \bibinfo {author}
  {\bibfnamefont {N.~N.}\ \bibnamefont {Iossad}}, \bibinfo {author}
  {\bibfnamefont {G.}~\bibnamefont {Rastelli}}, \bibinfo {author}
  {\bibfnamefont {S.}~\bibnamefont {Ciuchi}},\ and\ \bibinfo {author}
  {\bibfnamefont {A.~F.}\ \bibnamefont {Morpurgo}},\ }\href
  {https://doi.org/10.1038/nmat1774} {\bibfield  {journal} {\bibinfo  {journal}
  {Nature Materials}\ }\textbf {\bibinfo {volume} {5}},\ \bibinfo {pages} {982}
  (\bibinfo {year} {2006})}\BibitemShut {NoStop}%
\bibitem [{\citenamefont {Riley}\ \emph {et~al.}(2018)\citenamefont {Riley},
  \citenamefont {Caruso}, \citenamefont {Verdi}, \citenamefont {Duffy},
  \citenamefont {Watson}, \citenamefont {Bawden}, \citenamefont {Volckaert},
  \citenamefont {van~der Laan}, \citenamefont {Hesjedal}, \citenamefont
  {Hoesch}, \citenamefont {Giustino},\ and\ \citenamefont
  {King}}]{riley_crossover_2018}%
  \BibitemOpen
  \bibfield  {author} {\bibinfo {author} {\bibfnamefont {J.~M.}\ \bibnamefont
  {Riley}}, \bibinfo {author} {\bibfnamefont {F.}~\bibnamefont {Caruso}},
  \bibinfo {author} {\bibfnamefont {C.}~\bibnamefont {Verdi}}, \bibinfo
  {author} {\bibfnamefont {L.~B.}\ \bibnamefont {Duffy}}, \bibinfo {author}
  {\bibfnamefont {M.~D.}\ \bibnamefont {Watson}}, \bibinfo {author}
  {\bibfnamefont {L.}~\bibnamefont {Bawden}}, \bibinfo {author} {\bibfnamefont
  {K.}~\bibnamefont {Volckaert}}, \bibinfo {author} {\bibfnamefont
  {G.}~\bibnamefont {van~der Laan}}, \bibinfo {author} {\bibfnamefont
  {T.}~\bibnamefont {Hesjedal}}, \bibinfo {author} {\bibfnamefont
  {M.}~\bibnamefont {Hoesch}}, \bibinfo {author} {\bibfnamefont
  {F.}~\bibnamefont {Giustino}},\ and\ \bibinfo {author} {\bibfnamefont
  {P.~D.~C.}\ \bibnamefont {King}},\ }\bibfield  {journal} {\bibinfo  {journal}
  {Nature Communications}\ }\textbf {\bibinfo {volume} {9}},\ \href
  {https://doi.org/10.1038/s41467-018-04749-w} {10.1038/s41467-018-04749-w}
  (\bibinfo {year} {2018})\BibitemShut {NoStop}%
\bibitem [{\citenamefont {Hedin}(1965)}]{hedin_new_1965}%
  \BibitemOpen
  \bibfield  {author} {\bibinfo {author} {\bibfnamefont {L.}~\bibnamefont
  {Hedin}},\ }\href {https://doi.org/10.1103/PhysRev.139.A796} {\bibfield
  {journal} {\bibinfo  {journal} {Physical Review}\ }\textbf {\bibinfo {volume}
  {139}},\ \bibinfo {pages} {A796} (\bibinfo {year} {1965})}\BibitemShut
  {NoStop}%
\bibitem [{\citenamefont {Langreth}(1970)}]{langreth_singularities_1970}%
  \BibitemOpen
  \bibfield  {author} {\bibinfo {author} {\bibfnamefont {D.~C.}\ \bibnamefont
  {Langreth}},\ }\href {https://doi.org/10.1103/PhysRevB.1.471} {\bibfield
  {journal} {\bibinfo  {journal} {Physical Review B}\ }\textbf {\bibinfo
  {volume} {1}},\ \bibinfo {pages} {471} (\bibinfo {year} {1970})}\BibitemShut
  {NoStop}%
\bibitem [{\citenamefont {Heller}\ \emph {et~al.}(1979)\citenamefont {Heller},
  \citenamefont {Stechel},\ and\ \citenamefont
  {Davis}}]{heller_molecular_1979}%
  \BibitemOpen
  \bibfield  {author} {\bibinfo {author} {\bibfnamefont {E.~J.}\ \bibnamefont
  {Heller}}, \bibinfo {author} {\bibfnamefont {E.~B.}\ \bibnamefont
  {Stechel}},\ and\ \bibinfo {author} {\bibfnamefont {M.~J.}\ \bibnamefont
  {Davis}},\ }\bibfield  {journal} {\bibinfo  {journal} {The Journal of
  Chemical Physics}\ }\textbf {\bibinfo {volume} {71}},\ \href
  {https://doi.org/10.1063/1.438262} {10.1063/1.438262} (\bibinfo {year}
  {1979})\BibitemShut {NoStop}%
\bibitem [{\citenamefont {Heller}\ \emph {et~al.}(1980)\citenamefont {Heller},
  \citenamefont {Stechel},\ and\ \citenamefont
  {Davis}}]{heller_molecular_1980}%
  \BibitemOpen
  \bibfield  {author} {\bibinfo {author} {\bibfnamefont {E.~J.}\ \bibnamefont
  {Heller}}, \bibinfo {author} {\bibfnamefont {E.~B.}\ \bibnamefont
  {Stechel}},\ and\ \bibinfo {author} {\bibfnamefont {M.~J.}\ \bibnamefont
  {Davis}},\ }\bibfield  {journal} {\bibinfo  {journal} {The Journal of
  Chemical Physics}\ }\textbf {\bibinfo {volume} {73}},\ \href
  {https://doi.org/10.1063/1.440005} {10.1063/1.440005} (\bibinfo {year}
  {1980})\BibitemShut {NoStop}%
\bibitem [{\citenamefont {Ranninger}\ and\ \citenamefont
  {Thibblin}(1992)}]{ranninger_two-site_1992}%
  \BibitemOpen
  \bibfield  {author} {\bibinfo {author} {\bibfnamefont {J.}~\bibnamefont
  {Ranninger}}\ and\ \bibinfo {author} {\bibfnamefont {U.}~\bibnamefont
  {Thibblin}},\ }\bibfield  {journal} {\bibinfo  {journal} {Physical Review B}\
  }\textbf {\bibinfo {volume} {45}},\ \href
  {https://doi.org/10.1103/PhysRevB.45.7730} {10.1103/PhysRevB.45.7730}
  (\bibinfo {year} {1992})\BibitemShut {NoStop}%
\bibitem [{\citenamefont {Gunnarsson}\ \emph {et~al.}(1994)\citenamefont
  {Gunnarsson}, \citenamefont {Meden},\ and\ \citenamefont
  {Schönhammer}}]{gunnarsson_corrections_1994}%
  \BibitemOpen
  \bibfield  {author} {\bibinfo {author} {\bibfnamefont {O.}~\bibnamefont
  {Gunnarsson}}, \bibinfo {author} {\bibfnamefont {V.}~\bibnamefont {Meden}},\
  and\ \bibinfo {author} {\bibfnamefont {K.}~\bibnamefont {Schönhammer}},\
  }\href {https://doi.org/10.1103/PhysRevB.50.10462} {\bibfield  {journal}
  {\bibinfo  {journal} {Physical Review B}\ }\textbf {\bibinfo {volume} {50}},\
  \bibinfo {pages} {10462} (\bibinfo {year} {1994})}\BibitemShut {NoStop}%
\bibitem [{\citenamefont {Kas}\ \emph {et~al.}(2014)\citenamefont {Kas},
  \citenamefont {Rehr},\ and\ \citenamefont {Reining}}]{kas_cumulant_2014}%
  \BibitemOpen
  \bibfield  {author} {\bibinfo {author} {\bibfnamefont {J.~J.}\ \bibnamefont
  {Kas}}, \bibinfo {author} {\bibfnamefont {J.~J.}\ \bibnamefont {Rehr}},\ and\
  \bibinfo {author} {\bibfnamefont {L.}~\bibnamefont {Reining}},\ }\href
  {https://doi.org/10.1103/PhysRevB.90.085112} {\bibfield  {journal} {\bibinfo
  {journal} {Physical Review B}\ }\textbf {\bibinfo {volume} {90}},\ \bibinfo
  {pages} {085112} (\bibinfo {year} {2014})}\BibitemShut {NoStop}%
\bibitem [{\citenamefont {Aryasetiawan}\ \emph {et~al.}(1996)\citenamefont
  {Aryasetiawan}, \citenamefont {Hedin},\ and\ \citenamefont
  {Karlsson}}]{aryasetiawan_multiple_1996}%
  \BibitemOpen
  \bibfield  {author} {\bibinfo {author} {\bibfnamefont {F.}~\bibnamefont
  {Aryasetiawan}}, \bibinfo {author} {\bibfnamefont {L.}~\bibnamefont
  {Hedin}},\ and\ \bibinfo {author} {\bibfnamefont {K.}~\bibnamefont
  {Karlsson}},\ }\href {https://doi.org/10.1103/PhysRevLett.77.2268} {\bibfield
   {journal} {\bibinfo  {journal} {Physical Review Letters}\ }\textbf {\bibinfo
  {volume} {77}},\ \bibinfo {pages} {2268} (\bibinfo {year}
  {1996})}\BibitemShut {NoStop}%
\bibitem [{\citenamefont {Gumhalter}\ \emph {et~al.}(2016)\citenamefont
  {Gumhalter}, \citenamefont {Kovač}, \citenamefont {Caruso}, \citenamefont
  {Lambert},\ and\ \citenamefont {Giustino}}]{gumhalter_combined_2016}%
  \BibitemOpen
  \bibfield  {author} {\bibinfo {author} {\bibfnamefont {B.}~\bibnamefont
  {Gumhalter}}, \bibinfo {author} {\bibfnamefont {V.}~\bibnamefont {Kovač}},
  \bibinfo {author} {\bibfnamefont {F.}~\bibnamefont {Caruso}}, \bibinfo
  {author} {\bibfnamefont {H.}~\bibnamefont {Lambert}},\ and\ \bibinfo {author}
  {\bibfnamefont {F.}~\bibnamefont {Giustino}},\ }\href
  {https://doi.org/10.1103/PhysRevB.94.035103} {\bibfield  {journal} {\bibinfo
  {journal} {Physical Review B}\ }\textbf {\bibinfo {volume} {94}},\ \bibinfo
  {pages} {035103} (\bibinfo {year} {2016})}\BibitemShut {NoStop}%
\bibitem [{\citenamefont {Caruso}\ \emph {et~al.}(2015)\citenamefont {Caruso},
  \citenamefont {Lambert},\ and\ \citenamefont {Giustino}}]{caruso_band_2015}%
  \BibitemOpen
  \bibfield  {author} {\bibinfo {author} {\bibfnamefont {F.}~\bibnamefont
  {Caruso}}, \bibinfo {author} {\bibfnamefont {H.}~\bibnamefont {Lambert}},\
  and\ \bibinfo {author} {\bibfnamefont {F.}~\bibnamefont {Giustino}},\
  }\bibfield  {journal} {\bibinfo  {journal} {Physical Review Letters}\
  }\textbf {\bibinfo {volume} {114}},\ \href
  {https://doi.org/10.1103/PhysRevLett.114.146404}
  {10.1103/PhysRevLett.114.146404} (\bibinfo {year} {2015})\BibitemShut
  {NoStop}%
\bibitem [{\citenamefont {Lischner}\ \emph {et~al.}(2013)\citenamefont
  {Lischner}, \citenamefont {Vigil-Fowler},\ and\ \citenamefont
  {Louie}}]{lischner_physical_2013}%
  \BibitemOpen
  \bibfield  {author} {\bibinfo {author} {\bibfnamefont {J.}~\bibnamefont
  {Lischner}}, \bibinfo {author} {\bibfnamefont {D.}~\bibnamefont
  {Vigil-Fowler}},\ and\ \bibinfo {author} {\bibfnamefont {S.~G.}\ \bibnamefont
  {Louie}},\ }\bibfield  {journal} {\bibinfo  {journal} {Physical Review
  Letters}\ }\textbf {\bibinfo {volume} {110}},\ \href
  {https://doi.org/10.1103/PhysRevLett.110.146801}
  {10.1103/PhysRevLett.110.146801} (\bibinfo {year} {2013})\BibitemShut
  {NoStop}%
\bibitem [{\citenamefont {Guzzo}\ \emph {et~al.}(2014)\citenamefont {Guzzo},
  \citenamefont {Kas}, \citenamefont {Sponza}, \citenamefont {Giorgetti},
  \citenamefont {Sottile}, \citenamefont {Pierucci}, \citenamefont {Silly},
  \citenamefont {Sirotti}, \citenamefont {Rehr},\ and\ \citenamefont
  {Reining}}]{guzzo_multiple_2014}%
  \BibitemOpen
  \bibfield  {author} {\bibinfo {author} {\bibfnamefont {M.}~\bibnamefont
  {Guzzo}}, \bibinfo {author} {\bibfnamefont {J.~J.}\ \bibnamefont {Kas}},
  \bibinfo {author} {\bibfnamefont {L.}~\bibnamefont {Sponza}}, \bibinfo
  {author} {\bibfnamefont {C.}~\bibnamefont {Giorgetti}}, \bibinfo {author}
  {\bibfnamefont {F.}~\bibnamefont {Sottile}}, \bibinfo {author} {\bibfnamefont
  {D.}~\bibnamefont {Pierucci}}, \bibinfo {author} {\bibfnamefont {M.~G.}\
  \bibnamefont {Silly}}, \bibinfo {author} {\bibfnamefont {F.}~\bibnamefont
  {Sirotti}}, \bibinfo {author} {\bibfnamefont {J.~J.}\ \bibnamefont {Rehr}},\
  and\ \bibinfo {author} {\bibfnamefont {L.}~\bibnamefont {Reining}},\
  }\bibfield  {journal} {\bibinfo  {journal} {Physical Review B}\ }\textbf
  {\bibinfo {volume} {89}},\ \href {https://doi.org/10.1103/PhysRevB.89.085425}
  {10.1103/PhysRevB.89.085425} (\bibinfo {year} {2014})\BibitemShut {NoStop}%
\bibitem [{\citenamefont {Offi}\ \emph {et~al.}(2007)\citenamefont {Offi},
  \citenamefont {Werner}, \citenamefont {Sacchi}, \citenamefont {Torelli},
  \citenamefont {Cautero}, \citenamefont {Cautero}, \citenamefont {Fondacaro},
  \citenamefont {Huotari}, \citenamefont {Monaco}, \citenamefont {Paolicelli},
  \citenamefont {Smekal}, \citenamefont {Stefani},\ and\ \citenamefont
  {Panaccione}}]{offi_comparison_2007}%
  \BibitemOpen
  \bibfield  {author} {\bibinfo {author} {\bibfnamefont {F.}~\bibnamefont
  {Offi}}, \bibinfo {author} {\bibfnamefont {W.~S.~M.}\ \bibnamefont {Werner}},
  \bibinfo {author} {\bibfnamefont {M.}~\bibnamefont {Sacchi}}, \bibinfo
  {author} {\bibfnamefont {P.}~\bibnamefont {Torelli}}, \bibinfo {author}
  {\bibfnamefont {M.}~\bibnamefont {Cautero}}, \bibinfo {author} {\bibfnamefont
  {G.}~\bibnamefont {Cautero}}, \bibinfo {author} {\bibfnamefont
  {A.}~\bibnamefont {Fondacaro}}, \bibinfo {author} {\bibfnamefont
  {S.}~\bibnamefont {Huotari}}, \bibinfo {author} {\bibfnamefont
  {G.}~\bibnamefont {Monaco}}, \bibinfo {author} {\bibfnamefont
  {G.}~\bibnamefont {Paolicelli}}, \bibinfo {author} {\bibfnamefont
  {W.}~\bibnamefont {Smekal}}, \bibinfo {author} {\bibfnamefont
  {G.}~\bibnamefont {Stefani}},\ and\ \bibinfo {author} {\bibfnamefont
  {G.}~\bibnamefont {Panaccione}},\ }\bibfield  {journal} {\bibinfo  {journal}
  {Physical Review B}\ }\textbf {\bibinfo {volume} {76}},\ \href
  {https://doi.org/10.1103/PhysRevB.76.085422} {10.1103/PhysRevB.76.085422}
  (\bibinfo {year} {2007})\BibitemShut {NoStop}%
\end{thebibliography}%


\begin{thebibliography}{}

\bibitem[Goodvin et~al., 2006]{goodvin_greens_2006}
Goodvin, G.~L., Berciu, M., and Sawatzky, G.~A. (2006).
\newblock Green's function of the {Holstein} polaron.
\newblock {\em Physical Review B}, 74(24).

\bibitem[Gunnarsson et~al., 1994]{gunnarsson_corrections_1994}
Gunnarsson, O., Meden, V., and Schönhammer, K. (1994).
\newblock Corrections to {Migdal}'s theorem for spectral functions: {A}
  cumulant treatment of the time-dependent {Green}'s function.
\newblock {\em Physical Review B}, 50(15):10462--10473.

\bibitem[Kas et~al., 2014]{kas_cumulant_2014}
Kas, J.~J., Rehr, J.~J., and Reining, L. (2014).
\newblock Cumulant expansion of the retarded one-electron {Green} function.
\newblock {\em Physical Review B}, 90(8):085112.

\bibitem[Zhou et~al., 2018]{zhou_cumulant_2018}
Zhou, J.~S., Gatti, M., Kas, J.~J., Rehr, J.~J., and Reining, L. (2018).
\newblock Cumulant {Green}'s function calculations of plasmon satellites in
  bulk sodium: {Influence} of screening and the crystal environment.
\newblock {\em Physical Review B}, 97(3).

\end{thebibliography}

\end{document}

\title{Supplemental Material:}% Force line breaks with \\

\author{Bipul Pandey, Peter B. Littlewood}

\date{\today}% It is always \today, today,
             %  but any date may be explicitly specified
\maketitle

\section{Electron and Boson Green's Function}
In our single electron two site Holstein problem, we look at the electron addition spectra. For this problem, the ground state is the fock vacuum $|0\rangle$ which doesn't have any fermion or boson in it. We will now define the electron and boson Green's function for this problem with $|0\rangle$ as the ground state.

\textbf{The electron Green's function: } In retarded-time formalism \cite{kas_cumulant_2014} for a two orbitals ($n =\pm$) system described by \eqref{Holstein}, with $\{,\}$ as the anti-commutator, $c_n^\dagger /c_n$ as the electron creation/annihilation operators, and ${|0\rangle}$ as the fock vacuum, the RT one-particle electron addition Green's function $G(n;t)$ is the probability amplitude for a particle injected into orbital 'n' to be in 'n' after time t \cite{goodvin_greens_2006}.
\begin{equation}
\begin{aligned}
    G(n;t) &= -i \theta(t)\left\langle 0|\{c_n(t) ,{c_n^{\dagger}}\}|0\right\rangle\\
    &= -i \theta(t)\left\langle 0|\{e^{iHt}c_n e^{-iHt} ,{c_n^{\dagger}}\}|0\right\rangle
\end{aligned}
\label{Greens def}
\end{equation}
 The non-interacting($g=0$) or bare electron Green's function $G_o(\pm,t)$, given the bare energy eigenvalues $\varepsilon_{\pm}$ of  $H_o$ and evolution time 't', is as follows;
\begin{equation}
\begin{aligned}
    G_o(\pm,t) &=-i \theta(t)\left\langle 0|\{e^{iH_ot}c_n e^{-iH_ot} ,{c_n^{\dagger}}\}|0\right\rangle\\
    &=-i\theta(t) e^{-i\varepsilon_\pm t} 
\end{aligned}
    \label{Bare electron greens}
\end{equation} 

\textbf{The Boson Green's function: }
In RT formalism, with $[ , ]$ as the commutator, $b_N^\dagger/b_N$ as the boson creation/annihilation operators, ${|0\rangle}$ as the fock vacuum, the RT one-particle boson addition Green's function $D(N=\pm;t)$ is the probability amplitude for a 'N' type boson to remain in 'N' type after time t
\begin{equation}
\begin{aligned}
    \mathcal{D}(N,t) &= -i \theta(t)\langle 0|[b_N(t) ,{b_N^{\dagger}}]|0\rangle\\
     &= -i \theta(t)\langle 0|[e^{iHt}b_N e^{-iHt} ,{b_N^{\dagger}}]|0\rangle
\end{aligned}
\label{Boson greens}
\end{equation}
The non-interacting($g=0$) or bare boson operator for dispersionless bosons of frequency $\omega_o$ is given by;
\begin{equation}
\begin{aligned}
    \mathcal{D}(\pm,t) &=-i \theta(t)\langle 0|[e^{iH_\pm t}b_N e^{-iH_\pm t} ,{b_N^{\dagger}}]|0\rangle\\
    &=-i \theta(t)e^{- i\omega_o t}
\end{aligned}
    \label{Bare boson greens}
\end{equation}

\section{Spectral Function and Improper convergence of Delta Function}
In our work, the photo-emission spectral function $A(m,n;\omega)$ evaluated on the frequency axis is defined as;
 \begin{equation*}
     A(m,n;\omega) = \frac{1}{\pi} |\text{Im}G(m,n;\omega)|
 \end{equation*}
 This absolute valued definition of spectral function differs from the traditional definition and is necessary in numerical application because of the finiteness of the time axis. We explain this further in this section. The retarded time bare electron Green's function in frequency space is defined as;
\begin{equation}
\begin{aligned}
    G_o(k,\omega) &= \lim_{\eta\rightarrow0^+}\frac{1}{\omega -\varepsilon_k +i\eta}\\
    & =\mathscr{P}\Big[\frac{1}{\omega-\varepsilon_k} \Big] - i\pi\delta(\omega-\varepsilon_k)
\end{aligned}
\end{equation}
Here, $\mathscr{P}$ represents the principal value of the function it is acting on. We see that the imaginary part of this $G_o(k,\omega)$ has the poles at the energy eigenvalues $\varepsilon_k$ of the non-interacting part of Hamiltonian. From this, the traditional definition of the spectral function emerges;
\begin{equation}
    A_o(k,\omega) = -\frac{1}{\pi} \text{Im}(G_o(k,\omega)
    \label{traditional}
\end{equation}
Assuming a smooth transition from non-interacting to interacting system, we can extend this expression's validity to define interacting system's spectral function;
\begin{equation}
    A(k,\omega) = -\frac{1}{\pi} \text{Im}(G(k,\omega)
\end{equation}

In the context of Dirac Delta function we often use the following relationship:
\begin{equation}
\begin{aligned}
    \lim_{\eta\rightarrow 0^+} \frac{1}{x\pm i \eta} &=\lim_{\eta\rightarrow 0^+} \frac{x}{x^2+ \eta^2} \quad\mp\quad\lim_{\eta\rightarrow 0^+}  i\pi \frac{\eta}{\pi(x^2 + \eta^2)}\\
    & =\qquad \,\,\mathscr{P}\Big[\frac{1}{x} \Big] \qquad\mp\qquad i\pi\delta(x)
\end{aligned}
\end{equation}

The delta function in the imaginary part originates from the limit-definition (Sokhotski-Plemelj Theorem or Kramers Kronig Relations) of the function in the line right above it and hence is an idealization when it comes to numerical implementation. This is because in numerical implementation, explicitly demanding that $\eta$ must go to zero only from the positive side of the number line (since we demand $\eta\rightarrow0^+$) for a continuous function (bare electron green's function) is notoriously difficult. On top of this, the negative side then requires a sign flip in the definition of the delta function. Now, we no longer have a unified definition of delta function but rather a piece wise definition. This is still manageable when we have a single delta function i.e bare electron Green's function in any one half of the real line. But when we use the bare electron Green's function to compute actual Green's function in symmetric time domain and convert back to frequency domain, we now notice that we need to enforce this piece-wise definition of Green's function at every given frequency point. Furthermore, since there is a cutoff ($t_{max}$) in time, this manifests as oscillations in the frequency space in the order of $t^{-1}_{max}$. We are now at an impasse. We need a large $t_{max}$ (ideally $t_{max}\rightarrow \infty$) to properly capture the Green's function decay. But $t_{max}$ needs to be some large finite value for numerical implementation which manifests as violent small energy oscillation. In order to bypass this and reproduce correct answer for non-interacting as well as interacting fermionic systems, we can redefine the limit-definition of the function with an absolute value as follows;
\begin{equation}
\begin{aligned}
    \lim_{\eta\rightarrow 0} \frac{1}{x\pm i \eta} &=\lim_{\eta\rightarrow 0} \frac{x}{x^2+ \eta^2} \quad\mp\quad\lim_{\eta\rightarrow 0}  i\pi\Big| \frac{\eta}{\pi(x^2 + \eta^2)}\Big|\\
    & =\qquad \,\,\mathscr{P}\Big[\frac{1}{x} \Big] \qquad\mp\qquad i\pi\delta(x)
\end{aligned}
\end{equation}

Doing so, we now get a consistent single definition of the delta function on both sides of the number line. This manifests in our definition of the spectral function.

\begin{equation}
    A(k,\omega) = \frac{1}{\pi}|G(k,\omega)|
    \label{absolute}
\end{equation}

\begin{figure}
\centering
\begin{subfigure}{.5\textwidth}
  \centering
  \includegraphics[width=.9\linewidth]{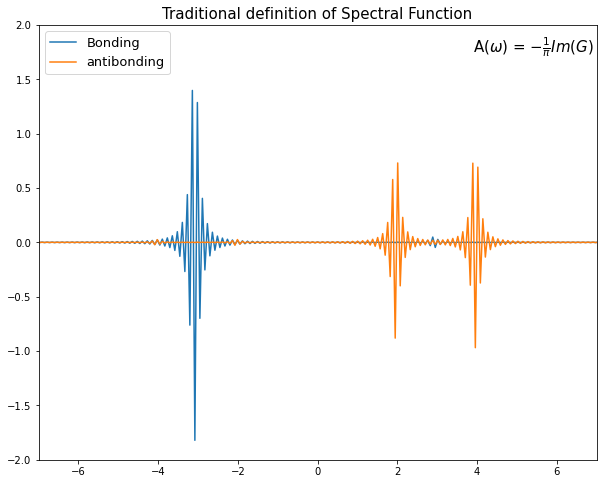}
  \caption{Traditional definition from equation \eqref{traditional}}
  \label{fig:sub1}
\end{subfigure}%
\begin{subfigure}{.5\textwidth}
  \centering
  \includegraphics[width=.9\linewidth]{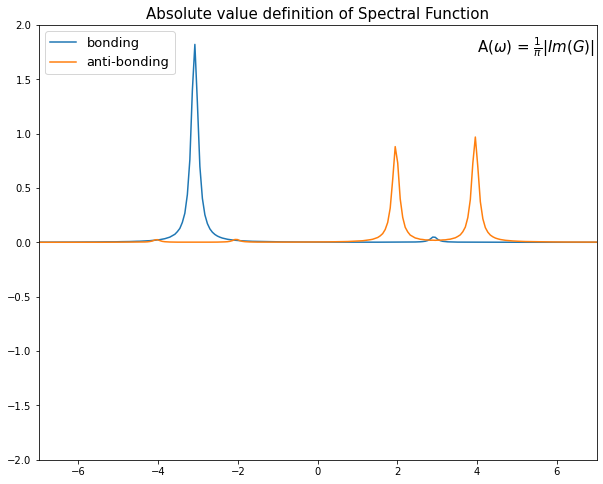}
  \caption{Absolute value definition from equation \eqref{absolute}}
  \label{fig:sub2}
\end{subfigure}
\caption{Two Definitions of Spectral Function for $g =1$, $\varepsilon_\pm= \mp 3$ and $\omega_o = 6$}
\label{fig:test}
\end{figure}

%%%%%%%%%%%%%%%%%%%%%%%%%%%%%%%%%%%%%%%%
\section{A curious case of the Holstein Hamiltonian}
The two site single electron Holstein Hamiltonian presented in the paper originates from a two site hopping model with site hopping completely determined by the hopping terms and not the bosons. Given the fermionic and bosonic creation/annihilation operators $c_i/c_i^\dagger$ and $b_i/b_i^\dagger$ for site $i = 1,2$, the Hamiltonian for such a system is \cite{gunnarsson_corrections_1994} ;

\begin{equation}
    H = \epsilon_o\sum_{i=1}^2 c_i^\dagger c_i + \omega_o \sum_{i=1}^2 b_i^\dagger b_i - t(c_1^\dagger c_2 + c_2^\dagger c_1)+ g_o\sum_{i=1}^2(b_i + b_i^\dagger)c_i^\dagger c_i
\end{equation}

A closer inspection of this Hamiltonian (last term) leads to the conclusion that boson emission or absorption do not cause any site hopping. Furthermore, the two sites are equivalent in energy i.e both have an energy $\epsilon_o$ and there is no preference in hopping from one site to another because the hopping amplitude $t$ is same for hopping along both directions. We can now go to the bonding and anti-bonding orbital basis with a change of variable for both fermions and bosons.
\begin{equation}
\begin{aligned}
    c_\pm&= \frac{c_1 \pm c_2}{\sqrt{2}}\qquad\qquad b_\pm= \frac{b_1 \pm b_2}{\sqrt{2}}\\
\end{aligned}
\end{equation}
With this change of variable, the Hamiltonain transforms to the one in the paper;
\begin{equation}
\begin{aligned}
&H=  \sum_{i=\pm} \varepsilon_\pm c_i^\dagger c_i  +  \sum_{i=\pm}\omega_o b_i^\dagger b_i +  g(c_+^\dagger c_+ + c_-^\dagger c_-)(b_+^\dagger + b_+)  +g(c_+^\dagger c_- + c_-^\dagger c_+)(b_-^\dagger + b_-)\\
&\text{where},\quad \epsilon_\pm = \epsilon_o\mp t \quad\text{and}\quad g = \frac{g_o}{\sqrt{2}}\\
\end{aligned}
\label {Holstein}
\end{equation}
Once this stage is reached, we can separate $H$ into a piece without any bosons $H_o$, a piece with only $(+)$ bosons - $H_+$ and a piece with only $(-)$ bosons - $H_-$ as shown in the paper. Now, we have transformed our system into two orbital system with a gap of $\Delta =2t$ where the hopping is entirely controlled by the $(-)$ bosons.
In case of the dihydrogen cation ($H_2^+$), there literally are two sites and a single electron. In this context, we can talk about the bonding and the anti-bonding orbitals originating from the original hydrogen molecule. In this idealized molecular system, the bosons are be the vibrational mode of the nuclei which may or may not cause inter-orbital transition. In this case, we have only one such vibrational mode because of the the diatomic structure- namely, nuclear motion along the line joining the two nuclei which stretches and compresses the bond length. We can then partition this bosonic space into the bosons that do in fact cause such transitions ($(-)$ bosons) and the ones that do not ($(+)$ bosons). 

In crystalline systems, the story becomes more general. We can have optical phonons which can cause transitions and phonons which do not. In this case, we can incorporate both of these behaviors with proper couplings and bosonic frequency by defining different phonon frequency $\omega_\pm$ and coupling constants $g_\pm$ for different phonons species. 
\section{Recursive relation for corrections}

In the paper, we saw how we can self-consistently update the power series $\mathcal{P}$ to find better and better approximation for itself. The full Dyson's series along with the perturbative nature of $\mathcal{P}$ also gives rise to recursive relations between correction functions $C_k$. By expanding $\mathcal{P}$ on both sides and comparing terms of same order in $g^2$ for $m^{th}$ orbital, we get;
\begin{equation}
    \begin{aligned}
    &C_{k}(m,t) = -i\int\displaylimits_{0}^{t}dt_2 \int\displaylimits_{0}^{t_2}d\tau \Bigg[\!e^{i\varepsilon_m\tau}\, \Sigma_o(m,\tau)C_{k-1}(m,t_2) +\sum_{n\neq m}\sum_{l=0}^{k}  e^{i\varepsilon_{m}\tau}\,\Sigma_o(n,\tau)\,C_{l}(n,\tau)\,C_{k-1-l}(m,t_2-\tau)\Bigg]\\
    \end{aligned}
    \label{recursive equation}
\end{equation}
Here too, inside the bracket, the first term is the self correction and the second term is the inter-band correction due to the effect of a different orbital $'n'$. By construction, we start with $C_o = 1$ for all bands. This scheme is useful for analytical proofs but cumbersome for numerical implementation. 

%%%%%%%%%%%%%%%%%%%%%%%%%%%%%%%%%%%%%%%%

\section{Derivation of Time-ordered Cumulant from Power Series}
The time-ordered cumulant is named so because of the use of time-ordered Green's function formalism. In this formalism, the electron lives in the $t>0$ branch of the Green's function while the hole lives in the $t<0$ branch of the Green's function. Therefore, there is no interaction between electrons and holes i.e both electrons and holes only talk amongst their own species.  Furthermore, the derivation was done with the assumption that in the Dyson's equation, any $n^{th}$ orbital's electron Green's function $G(n,t)$ depends only on $n^{th}$ orbital self energy $\Sigma(n,t)$ and not the total self energy $\Sigma(t)$ when it is evolving in time. This would be true if we knew the actual approximation free self-energy for the $n^{th}$ orbital. But in every practical case, what we have is some truncated self energy that neglects the boson mediated inter-band scattering effect mediated. Hence, using some approximate $\Sigma(n,t)$ instead of power series corrected $\Sigma(t)$ in Green's function evolution isolates each orbital as a core-hole problem. In multi-band system, this is an even more stringent condition because each orbital only scatters to itself regardless of it being a hole state or an electron state or there being other electron or hole states around. 

In the context of our single electron two orbital problem, this means that the hole and the electron states should be treated independent of each other and hence $H_-$ is neglected from the total Hamiltonian. Therefore, there is no inter-band correction term ($P_{IC} =0$) and all the dynamics is governed by the self correction term. The corrected self energy for electron/hole ($e/h$)for this case is;
\begin{equation*}
\begin{aligned}
        \Sigma^{e}(t) &=g^2 \Sigma_o^{e}(t)\mathcal{P}_e(t)\, =\,  -i\theta(t)g^2 \big[ - e^{-i(\varepsilon_e +\omega_o)t}\big] \mathcal{P}_e(t)\\
        \Sigma^{h}(t) &= g^2 \Sigma_o^{h}(t)\mathcal{P}_h(t)\, =\,  i\theta(t)g^2 \big[ - e^{-i(\varepsilon_h +\omega_o)t}\big] \mathcal{P}_h(t)
\end{aligned}
\end{equation*}

Since there is no inter-band scattering correction in power series correction equation for both electrons and holes, the sets of equation decouple. For electron, the correction equation is as follows;
\begin{equation*}
    \begin{aligned}
        \mathcal{P}_e(t)=1+ \Big[-ig^2\!\!\int\displaylimits_{0}^{t} dt_2 \int\displaylimits_{0}^{t_2}\!\!d{\tau}\, e^{i\varepsilon_e\tau}\Sigma_o(\tau)\mathcal{P}_e(t_2)\Big]
    \end{aligned}
    \label{single-disp-boson}
\end{equation*}

Expanding power series $\mathcal{P}_e$ on both sides and comparing terms of same order in $g^2$ across the equality, we generate the following higher order corrections.
\begin{equation*}
    C_1(t) =  \Big[\frac{e^{-i\omega_o t} + i\omega_o t -1}{\omega_o^2}\Big] \quad \text{and}\quad C_k (t) = \frac{C_1(t)^k}{k!}
\end{equation*}
Summing all of these corrections gives us the exact result for the core hole problem.
\begin{equation}
G^e(t) = G_o^e(t)\mathcal{P}_e(t) = G_o^e(t)\,e^{g^2 C_1(t)}
\label{core-hole-cumulant}
\end{equation}
An equivalent derivation for the hole cumulant can be performed by following the steps outlined above.

\section{Derivation of Retarded-time Cumulant from Power Series}
In the cited papers, the authors derive cumulant results for $H_o + H_-$ rather than $H$ because the effect of $H_+$ is like that of core-hole problem in that it causes no inter-band transition. Here we choose to use this same model for proper comparison with the literature \cite{zhou_cumulant_2018}.
For bosons of frequency $\omega_o$ and two bands with bare energies $\varepsilon_{+}$ and $\varepsilon_{-}$, if the above assumptions about explicit band independence of corrections hold true, we can compute the correction series exactly. The bare band retarded self energies are \cite{zhou_cumulant_2018, gunnarsson_corrections_1994};
\begin{equation*}
\begin{aligned}
    \Sigma_o(+,\omega) &=  \Big(\frac{1}{\omega - \varepsilon_{+} - \omega_o -i \eta}\Big)\\
    \Sigma_o(-,\omega) &=  \Big(\frac{1}{\omega - \varepsilon_{-} - \omega_o -i \eta}\Big) 
\end{aligned}
\end{equation*}
In the literature \cite{zhou_cumulant_2018}, the authors choose to write the total self energy without power series correction.
\begin{equation}
    \Sigma(\omega) = g^2 \Sigma(+,\omega) + g^2 \Sigma(-,\omega)
\end{equation}
We choose to correct the total self energy with power series correction inside as shown in our paper. The total self energy $\Sigma(t)$ for such a system given each level's self energies $\Sigma(m,t)$ is then;
\begin{equation*}
    \Sigma(t) = g^2\Big(\sum_{m=\pm} \Sigma_o(m,t)\Big)\mathcal{P}(t)
\end{equation*}
 Here, both the terms originating from different boson species look identical because of the symmetry of the problem (i.e $\omega_o$ and $g$ being same in both species). We could easily change $\omega_o$ to $\omega_\pm$ between the two boson types and repeat this analysis. If there are are two coupling constants $g_\pm$ for each boson species $(\pm)$, then the self energy can be written in terms of a third dummy coupling constant $g$ as;
 \begin{equation*}
    \Sigma(t) = \Big(\sum_{m=\pm} g_m^2 \Sigma_o(m,t)\Big)\mathcal{P}(t) = g^2\Big(\sum_{m=\pm} \Big[\frac{g_m}{g}\Big]^2 \Sigma_o(m,t)\Big)\mathcal{P}(t) 
\end{equation*}
We then include $g_m$ in the bare self energy $\Sigma_o(m,t)$ and expand the power series in terms of $g^2$ instead of $g_m^2$. In the end, we set this $g$ to be 1. 

 As mentioned in our work, for retarded cumulant derivation, we assume that since the orbital energy gap $\Delta$ is much smaller than the boson frequency $\omega_o$, we can assume that the power series corrections are explicitly orbital independent. This greatly simplifies our power series equation because we can use the temporal contraction relation between the power series pieces without caring about the orbital index. Coming back to the problem at hand with same $\omega_o$ and $g$, we can then write down the recursion relation for the correction power series for the $n^{th}$ band after using the contraction relation as ;

\begin{equation*}
\begin{aligned}
        \mathcal{P}(t) = 1+(-ig^2) &\sum_{m=\pm} \int\displaylimits_{0}^{t} \!\!d t_2\!\!\!\int\displaylimits_{0}^{t_2}d\tau e^{i \varepsilon_{n}\tau} \Sigma_o(m,\tau) \mathcal{P}(\tau)\\
\end{aligned}
\end{equation*}
If we solve the above equation for correction for the first orbital $\varepsilon_+$ with these band self energies, we get  the retarded cumulant expressions;
\begin{equation}
\begin{aligned}
\mathcal{P}(t) &= e^{g^2[C_1^+(t)]}\\
C_1^+(t) &= \Big(\frac{e^{-i\omega_o t} + i\omega_o t -1}{\omega_o^2}\Big) + \Big(\frac{e^{-i{\Tilde\omega_o} t} + i{\Tilde\omega_o}t -1}{{\Tilde\omega_o}^2}\Big)\\
\text{where,\,}&{\Tilde\omega_o} = \omega_o +(\varepsilon_{-}-\varepsilon_{+}) = \omega_o +\Delta\\
\end{aligned}
\end{equation}

Here, we see two distinct terms in the cumulant $C_1^+$. The first term generates satellites at intervals of $\omega_o$ from $\varepsilon_+$ orbital. This is a satellite generated by the electron interacting with a (-) boson and jumping down to $\varepsilon_+$ orbital from $\varepsilon_+$ orbital. The second term generates the satellite at intervals of $\omega_o+\Delta$ due to the electron interacting with a (-) boson and jumping up form $\varepsilon_+$ to $\varepsilon_-$ orbital. So the satellites now appear from the final orbital rather than the initial orbital.  In reality however, the satellites due to $(-)$ plasmon from one orbital should emerge in the interval of $\omega_o$ and not $\omega_o + \Delta$ from the final orbital. So, the retarded cumulant is getting only the very first ${\Tilde{\omega_o}}$ satellite correct. Fortunately, in the limit of $\Delta <<\omega_o$, these ${\Tilde{\omega_o}}$ satellites are so far off from the quasiparticle that they don't modify the quasiparticle spectra appreciably. And hence, the explicit orbital independence assumption becomes valid.

For sanity check, we can compute the power series numerically. We see that the $20^{th}$ order Power series converges to the spectral function given by the retarded cumulant expression in the literature- here referred to as "Cumulant corrected". Any further attempt to update the power series just results in the same function which means that we have converged to the exact solution.
\begin{figure}[htp]
    \centering
    \includegraphics[width = 0.7\textwidth]{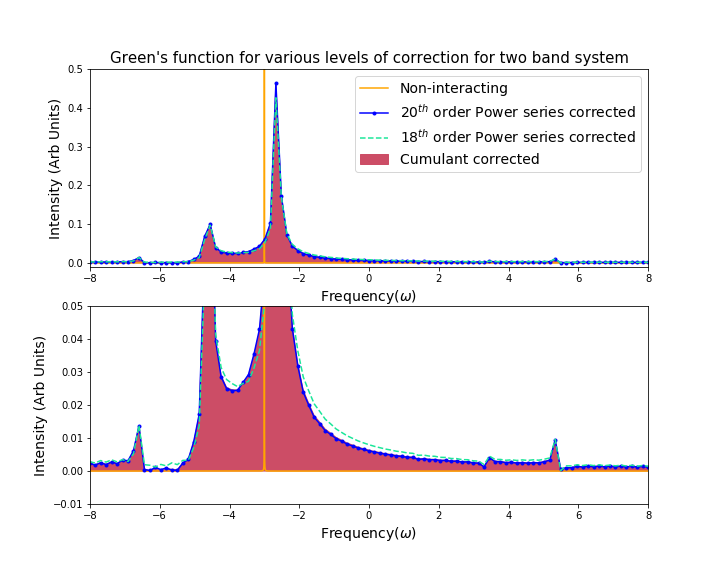}
    \caption{Numerically computed $20^{th}$ order retarded cumulant solution converges exactly to the predicted expression for retarded cumulant. Any more power series update just results in the same function. }
    \label{retarded cumulant}
\end{figure}

\section{Details of Exact Diagonalization}
In this section, we will briefly outline the construction of the two orbital Holstein Hamiltonian and the process of exact diagonalization. For a system with a single electron, N number of (+) bosons and N number of (-) bosons, there are three different components in the wave function - one for the electron and two for the two different bosons. For single electron, the electron wave function has three distinct entries each of which can either be 0 or 1. This is because of Pauli exclusion principle.
\begin{equation}
\begin{aligned}
    \ket{\psi_e} &=\ket{n_v, n_+, n_-} \qquad \text{where,}\\
    n_v &= \text{vacuum designator}\\
    n_+ &= \text{+ orbital designator}\\
    n_- &= \text{- orbital designator}\\
\end{aligned}
\end{equation}

Here, if there are no electrons in the system $n_v =1$ denoting electron vacuum. Presence of any electron in the system implies that $n_v =0$. If the electron is in $+$ orbital, $n_+ =1$ and otherwise $n_+=0$. Similarly, if the electron is in $-$ orbital, $n_- =1$ and otherwise $n_-=0$. For bosons, there is no restriction on the number of bosons that can coexist at a time. But for the sake of exact diagonalization, we need to enforce a cutoff that the maximum possible boson number is N in order to cutoff the Hamiltonian- the idea being that as $N\rightarrow\infty$, this finite Hamiltonian's eigenvalues approaches the exact eigenvalues. Any given $m^{th}$ wave function denoting that there are "m" bosons in the system for the ($\pm$) boson is as follows;
\begin{equation}
\begin{aligned}
    \ket{\Phi_\pm} &= \ket{n_0,n_1, n_2, n_3,..,n_m,..,n_{N-1},n_N} \quad\text{where,}\\
    n_m &= 1\\
    n_{k\neq m} &=0
\end{aligned}
\end{equation}
Here $n_0=1$ indicates boson vacuum. Since our boson wave function is based on the boson number rather than states, at any given time, only one of the $n_i$ can be non-zero. For instance, if there are two (+) bosons, only $n_2=1$ and all other $n_{i\neq 2} = 0$. For the entire single electron two plasmon bath system, any total wave function is then;
\begin{equation}
    \ket{\Psi(a,b,c)} = \ket{\psi_e^a} \bigotimes \ket{\Phi_+^b} \bigotimes \ket{\Phi_-^c}
\end{equation}
Here, $0\leq b,c \leq N$ by construction. In this system, there are $3(N+1)^2$ basis vectors. Because the Hamiltonian matrix scales as $(N+1)^2$,computation becomes exceedingly expensive with increasing boson number. 
\begin{figure}
    \centering
    \includegraphics[width = 0.7\textwidth]{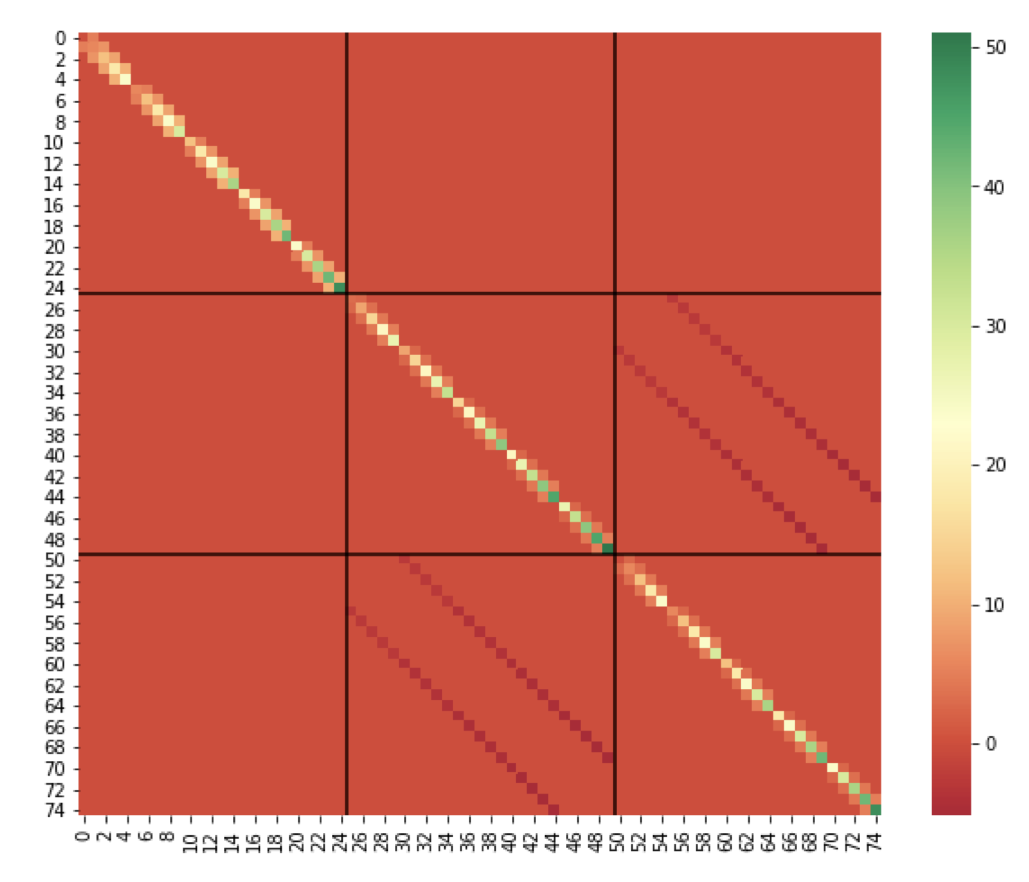}
    \caption{Hamiltonian with maximum of 4 plasmons
    (bosons) in each plasmon species ($\pm$) is already a $75\times75$ matrix}
    \label{4 boson H}
\end{figure}

Once, we construct this Hamiltonian, we can then find the eigenvalues and eigenvectors for it. The eigenvalue-eigenvector pair is represented as $\{\varepsilon_i, \ket{i}\}_i$ and there are $3(N+1)^2$ of them. The choice of boson number is dependent on the energy scale that we are looking at. With increasing $N$, we get the ability to resolve events closer in energy at the expense of computation time. At large plasma frequency, events happen far apart from each other and hence we only need a few bosons to resolve the system properly. At small plasma frequency however, since plasmonic shake offs are very close to each other, we need a large number of bosons to properly resolve such events.
\begin{figure}[H]
    \centering
    \includegraphics[width = 0.9\textwidth]{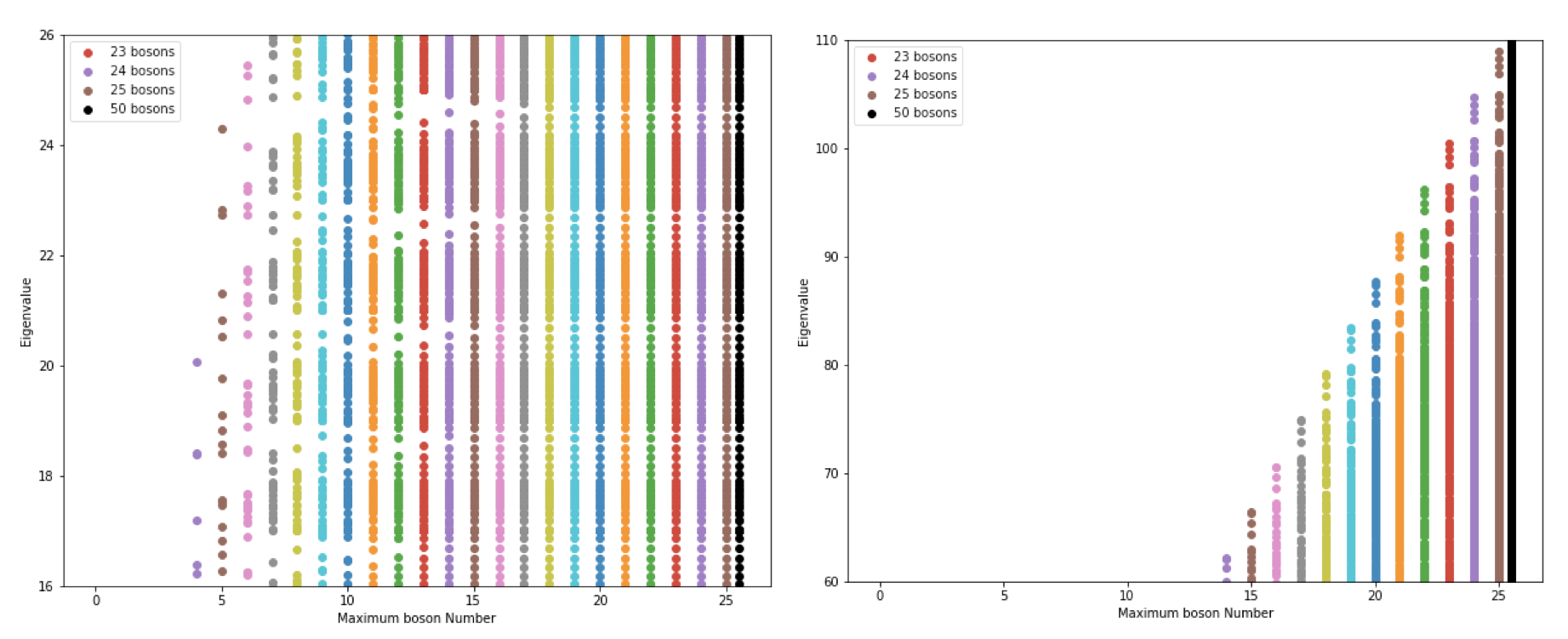}
    \caption{With increasing boson number, the eigenvalue differences between different sized system becomes smaller and smaller. The noticeable difference in above figures between different sized systems are only at large eigenvalues that lie in the eigenvalue continuum (right).}
    \label{Eigenvalue_spectra}
\end{figure}

Finally, we are in a position to compute the single particle Green's function. Here $\ket{0}$ implies total vacuum (Fermion as well as boson vacuums' outer product).

\begin{equation}
\begin{aligned}
    G(m,n;t) &= -i \theta(t)\left\langle 0|\{c_m(t) ,{c_n^{\dagger}}\}|0\right\rangle\\
    &= -i \theta(t)\left\langle 0|\{e^{iHt}c_m e^{-iHt} ,{c_n^{\dagger}}\}|0\right\rangle\\
    &= -i \theta(t)\sum_i\sum_j\left\langle 0|\{e^{iHt}\ket{i}\bra{i}c_m \ket{j}\bra{j}e^{-iHt} ,{c_n^{\dagger}}\}|0\right\rangle\\ 
\end{aligned}
\label{Greens definition}
\end{equation}
The only piece that survived in this Green's function after we open the anti-commutator is given by,
\begin{equation}
\begin{aligned}
    G(m,n;t)
    &= -i \theta(t)\sum_i\sum_j\left\langle 0|e^{iHt}\ket{i}\bra{i}c_m \ket{j}\bra{j}e^{-iHt} ,{c_n^{\dagger}}|0\right\rangle\\ 
    &= -i \theta(t)\sum_i\sum_j\left\langle 0|e^{i\varepsilon_it}\ket{i}\bra{i}c_m \ket{j}\bra{j}e^{-i\varepsilon_j t} ,{c_n^{\dagger}}|0\right\rangle\\ 
\end{aligned}
\label{Greens definition RT}
\end{equation}
This is the Green's function from exact diagonalization that we plot in our work.

\bibliography{Power_Series_5}